\documentclass[12pt,preprint]{aastex}
\usepackage[twocolumn]{emulateapj5}

\usepackage{graphicx}

\def\ergs{erg\,s$^{-1}$}
\def\ergscm2{erg\,s$^{-1}$cm$^{-2}$}
\def\ss{s\,s$^{-1}$}
\def\uu{4U\,0142+614\,}
\def\ee{1E\,1048-5937\,}
\def\kes{1E\,1841-045\,}
\def\axj{AX\,J1844-0258\,}
\def\rxs{1RXS\,J1708-4009\,}
\def\ea{1E\,2259+586\,}
\def\xte{XTE\,J1810-197\,}
\def\cxo{CXOU\,J0100-7211\,}
\def\wes{CXOU\,J1647-4552\,}
\def\1e{1E\,1547.0-5408\,}
\def\sgra{SGR\,1806-20\,}
\def\sgrb{SGR\,1900+14\,}
\def\sgrc{SGR\,0526-66\,}
\def\sgrd{SGR\,1627-41\,}
\def\sgre{SGR\,1801-23\,}

\newcommand{\XMM}{{\it XMM--Newton}\,}

\newcommand{\RXTE}{{\it R}XTE\,}
\newcommand{\INT}{{\it INTEGRAL}\,}
\newcommand{\CXO}{{\it Chandra}\,}
\newcommand{\Swift}{{\it Swift}}

\shorttitle{RCS in magnetars}
\shortauthors{Nanda Rea et al.}

\begin{document}

\title{Resonant cyclotron scattering in magnetars' emission}

\author{N. Rea\altaffilmark{1}, S. Zane\altaffilmark{2},
R. Turolla\altaffilmark{3,2}, M. Lyutikov\altaffilmark{4} and D. G\"otz\altaffilmark{5}}
\altaffiltext{1}{University of Amsterdam, Astronomical Institute ``Anton Pannekoek'', Kruislaan, 403, 1098~SJ, Amsterdam, The Netherlands}
\altaffiltext{2}{Mullard Space Science Laboratory, University College London, Holmbury St. Mary, Dorking, Surrey, RH5 6NT, UK}
\altaffiltext{3}{Department of Physics, University of Padova, Via Marzolo 8, I-35131
Padova, Italy}
\altaffiltext{4}{Department of Physics, Purdue University, 525 Northwestern Avenue, West Lafayette, IN 47907, USA}
\altaffiltext{5}{CEA Saclay, DSM/DAPNIA/Service d'Astrophysique, Gif sur Yvette, France}

\input psfig.sty

\begin{abstract}

We present a systematic fit of a model of resonant cyclotron
scattering (RCS) to the X-ray data of ten magnetars, including
canonical and transient anomalous X-ray pulsars (AXPs), and soft gamma
repeaters (SGRs). In this scenario, non-thermal magnetar spectra in
the soft X-rays (i.e. below $\sim 10$ keV) result from resonant
cyclotron scattering of the thermal surface emission by hot
magnetospheric plasma.  We find that this model can successfully
account for the soft X-ray emission of magnetars, while using the same
number of free parameters than the commonly used empirical blackbody
plus power-law model. However, while the RCS model can alone reproduce
the soft X-ray spectra of AXPs, the much harder spectra of SGRs below
10\,keV, requires the addition of a power-law component (the latter
being the same component responsible for their hard X-ray
emission). Although this model in its present form does not explain
the hard X-ray emission of a few of these sources, we took this
further component into account in our modeling not to overlook their
contribution in the $\sim$4-10\,keV band. We find that the entire
class of sources is characterized by magnetospheric plasma with a
density which, at resonant radius, is about 3 orders of magnitudes
higher than $n_{GJ}$, the Goldreich-Julian electron density. The
inferred values of the intervening hydrogen column densities, are also
in better agreement with more recent estimates inferred from the fit
of single X-ray edges. For the entire sample of observations, we find
indications for a correlation between the scattering depth and the
electron thermal velocity, and the field strength. Moreover, in most
transient anomalous X-ray pulsars the outburst state is characterized
by a relatively high surface temperature which cools down during the
decay, while the properties of the magnetospheric electrons vary in a
different way from source to source. Although the treatment of the
magnetospheric scattering used here is only approximated, its
successful application to all magnetars we considered shows that the
RCS model is capable to catch the main features of the spectra
observed below $\sim10$~keV.

\end{abstract}


\section{Introduction}
\label{intro}

The neutron star world, as we knew it until not long ago, appeared
mainly populated by radio pulsars (PSRs, about 2000 objects). In the
last two decades diverse, puzzling classes of isolated neutron stars
(NSs), with properties much at variance with those of canonical PSRs,
were discovered: the anomalous X-ray pulsars (AXPs), the soft gamma
repeaters (SGRs; Woods \& Thompson~2006; Mereghetti 2008), the
rotating radio transients (RRATs; McLaughlin et al.~2006), and the
X-ray dim isolated neutron stars (XDINSs; Haberl~2007).  Among these,
the AXPs and SGRs are, in some sense, the most peculiar, since they
are believed to host ultra-magnetized NSs, with a magnetic field
$\approx 10^{14}$--$10^{15}$\,G, in excess of the critical magnetic
field, $B_{crit} \equiv m_e^2c^3/(e\hbar)=4.414\times 10^{13}$~G, at
which the cyclotron energy equals the rest mass energy for an electron
(Duncan \& Thompson 1992; Thompson \& Duncan 1993, 1995, 1996).

The magnetar candidates (about fifteen known objects) are
characterized by slow X-ray pulsations ($P\sim2$--12\,s) and large
spin-down rates ($\dot P\sim 10^{-10}$--$10^{-12}$ \ss).  A
distinctive property is their high persistent X-ray luminosity
($L\approx 10^{34}$--$10^{36}$\ergs), which exceeds the spin-down
luminosity typically, by two orders of magnitude. Thus, magnetar X-ray
emission can not be explained in terms of rotational energy losses.
Measurements of spin periods and period derivatives, assuming that the
latter are due to electromagnetic dipolar losses, lend further support
to the idea that these objects contain neutron stars endowed with an
ultra-strong magnetic field.  Although the magnetar model has become
increasingly popular, alternative scenarios to explain the enigmatic
properties of these sources have been proposed. Among these, models
involving accretion from a fossil disk, formed in the supernova event
which gave birth to the neutron star, are still largely plausible
(e.g. van Paradijs et al.~1995; Chatterjee, Hernquist \& Narayan~2000;
Perna, Heyl \& Hernquist 2000).

Magnetar X-ray emission may be qualitatively separated into two
components, a low-energy, $\lesssim 10$\, keV, and a high-energy one,
$\gtrsim 20$\, keV. It is likely, although not proved yet, that
different emission mechanisms are responsible for the two
components. The low energy component is typically fit with either a
blackbody with a temperature $kT\sim 0.3-0.6$\, keV and a power-law
with a relatively steep photon index, $\Gamma\sim 2$--4, or two
blackbodies with $kT_1\sim 0.3$\,keV and $kT_2\sim 0.7$\,keV (see
Woods \& Thompson 2006 and Mereghetti 2008 for a review).  In a few
cases the low-energy component of SGR spectra has been fit with a
single power-law, but recent longer observations have shown that, also
for these sources, a blackbody component is required (Mereghetti et
al. 2005a). The high-energy component, discovered from four AXPs
(Kuiper et al. 2004, 2006) and two SGRs (Mereghetti et al. 2005b;
Molkov et al. 2005; G\"otz et al. 2006) has in general a quite hard
spectrum (modeled by a power-law), and accounts for about half of the
bolometric luminosity of these sources.  This makes it crucial to
consider in any spectral modeling the whole 1--200\,keV spectrum,
where $>90\%$ of the magnetar emission is concentrated, instead of
focussing on the soft X-ray range alone.  Furthermore, the discovery
of magnetar counterparts in the radio and infrared/optical bands
(Camilo et al. 2006; Hulleman et al. 2000) enforced the idea that
their multi-wavelength spectral energy distribution is by far more
complex than the simple superposition of blackbody (BB) and power-law
(PL) distributions.

The purpose of this paper is to provide a physical interpretation of
the soft X-ray component ($\lesssim 10$~keV) through a detailed
analysis of magnetar spectra. Our starting point is the work by
Thompson, Lyutikov \& Kulkarni (2002, TLK in the following), who
pointed out that resonant scattering in magnetar magnetospheres may
explain the non-thermal emission observed in magnetar candidates. Due
to the presence of hot plasma in the neutron star coronae, the thermal
emission from the neutron star surface/atmosphere gets distorted
through efficient resonant cyclotron scattering. Resonant cyclotron
scattering has been first studied in the accretion columns of neutron
star X-ray binary systems or in their atmospheres (Wasserman \&
Salpenter 1980; Nagel 1981; Lamb, Wang \& Wasserman 1990). Lyutikov \&
Gavriil (2006) computed, in an approximated and semi-analytical way,
the effect of multiple resonant scatterings of soft photons in the
magnetosphere, and found that the emerging spectrum is non-thermal,
with a shape that may resemble the observed blackbody plus
power-law. This model was preliminarily fit to the spectrum of the AXP
\ee\, (Lyutikov \& Gavriil~2006), although the magnetospheric
parameters were held fixed during the modeling.  Rea et al.~(2007a,b)
implemented in {\tt XSPEC} a more refined version in which also these
parameters are minimized during the fit (see
\S\ref{sec:xspec}), and successfully modeled a simultaneous \Swift\,
and \INT\, observation of \uu. In the following, we refer to this {\tt
XSPEC} model as the RCS model, where RCS stands for Resonant Cyclotron
Scattering. G\"uver et al.~(2007a,b) fit a similar model to two AXPs,
taking into account for the fact that the thermal emission from the
star surface is not a blackbody if the presence of an atmosphere is
accounted for (see also \S\ref{discussion}).  More detailed, fully 3D
Monte Carlo simulations of multiple resonant scattering in the star
magnetosphere have been very recently presented by Fernandez \&
Thompson (2007; see also Nobili, Turolla \& Zane 2008) but not
directly applied to the data yet (this will be done in a subsequent
paper).

In this paper we present a systematic application of the RCS model to
observations of all AXPs and SGRs. We consider the deepest X-ray
pointings available up to now for these sources, obtained making use
of the large throughput of the \XMM\, satellite. For a subset of
sources, which have been detected in the hard X-ray range, we also
consider a joint fit with the \INT\, spectra in order to study
systematically the relation between hard and soft X-rays production
mechanisms.

The paper is organized as follows. The basic concepts behind the RCS
model and its {\tt XSPEC} implementation are summarized in
\S~\ref{rcs}.  In \S\ref{obs} we report the observations and
the data analysis. Results of the spectral modeling
are presented in \S\ref{results}, and discussed in \S\ref{discussion}.
Conclusions follow.

\section{Resonant Cyclotron Scattering}
\label{rcs}

\subsection{The model}
\label{rcsmodel}

Before discussing our {\tt XSPEC} model and the implications of our
results, we briefly touch on some properties of the RCS model which
directly bear to the physical interpretation of the fitting parameters
and their comparison with similar parameters introduced in other
theoretical models. The basic idea follows the original suggestion by
TLK, who pointed out that a scattering plasma may be supplied to the
magnetosphere by plastic deformations of the crust, which twist the
external magnetic field and push electric currents into the
magnetosphere. The particle density of charge carries required to
support these currents may largely exceed the Goldreich-Julian charge
density (Goldreich \& Julian 1969). Furthermore, it is expected that
instabilities heat the plasma.

Following this idea, Lyutikov \& Gavriil (2006) studied how
magnetospheric plasma might distort the thermal X-ray emission
emerging from the star surface through efficient resonant cyclotron
scattering.  If a large volume of the neutron star magnetosphere is
filled by a hot plasma, the thermal (or quasi-thermal) cooling radiation
emerging from the star surface will experience repeated scatterings at
the cyclotron resonance.
The efficiency of the process is quantified by the scattering optical
depth, $\tau_{res}$,

\begin{equation}
\tau_{res}= \int \sigma_{res} n_e dl = \tau_0 ( 1+ \cos^2 \alpha)
\label{eqtaures}
\end{equation}
where

\begin{equation}
\sigma_{res} = {\sigma_T \over 4} { (1+ \cos ^2 \alpha) \omega^2 \over (
\omega - \omega_B)^2 +\Gamma^2/4}
\label{sigres}
\end{equation}
is the (non-relativistic) cross-section for electron scattering in the
magnetized regime, $n_e$ is the electrons number density, $\alpha$ is the
angle between the photon propagation
direction and the local magnetic field,
$\Gamma = 4 e^2 \omega_B^2 / 3 m_e c^3$ is the natural width of the first
cyclotron harmonic, $\sigma_T$ is the
Thomson scattering cross-section, and
\begin{equation}
\tau_0 = { \pi^2 e^2 n_e r\over 3 m_e c \omega_B} \, .
\label{tau0}
\end{equation}
Here $r$ is the radial distance from the center of the star,
$\omega_B = eB/m_{e}c$ is the electron cyclotron frequency, and $B$ is
the local value of the magnetic field. At energies corresponding to soft
X-ray photons, the resonant scattering optical depth greatly exceeds that
for Thomson scattering, $\tau_{T} \sim n_e \sigma_T r$,
\begin{equation}
 { \tau_{res} \over \tau_{T} } \sim { \pi \over 8} { m_ec^3 \over e^2
\omega_B} \sim 10^5 \left( \frac{1\ \mathrm{keV}}{\hbar \omega_B}
\right)\,.
\end{equation}
This implies that even a relatively small amount of plasma
present in the magnetosphere of the NS may considerably modify the
emergent spectrum.

The RCS model developed by Lyutikov \& Gavriil (2006), and used in
this investigation, is based on a simplified, 1D semi-analytical
treatment of resonant cyclotron up-scattering of soft thermal photons,
under the assumption that scattering occurs in a static,
non-relativistic, warm medium and neglecting electron recoil. The
latter condition requires $\hbar\omega\ll m_ec^2$.  Emission from the
neutron star surface is treated assuming a blackbody spectrum, and
that seed photons propagate in the radial direction.
Magnetospheric charges are taken to have a top-hat
velocity distribution centered at zero and extending up to
$\pm\beta_T$. Such a velocity distribution mimics a scenario in which
the electron motion is thermal (in 1D because charges stick to the
field lines). In this respect, $\beta_T$ is associated to the mean
particle energy and hence to the temperature of the 1D electron
plasma.
Since scatterings
with the magnetospheric electrons occur in a thin shell of width $H
\sim \beta_{T} r/3 \ll r$
around the ``scattering sphere'', one can treat
the scattering region as a plane-parallel slab. Radiation transport is
tackled by assuming that photons can only propagate along the slab
normal, i.e. either towards or away from the star. Therefore,
$\cos\alpha=\pm 1$ in eq.~(\ref{eqtaures}) and it is $\tau_{res}= 2
\tau_0$; the electron density is assumed to be constant through the
slab. We notice that the model does not account for the bulk
motion of the charges. This is expected since the starting point is
not a self-consistent calculation of the currents but a prescription
for the charge density.  As a consequence, the electron velocity and
the optical depth are independent parameters, although in a more
detailed treatment this might not be the case (Beloborodov \& Thompson
2007).

Although Thomson scattering conserves the photon energy in the
electron rest frame, the (thermal) motion of the charges induces a
frequency shift in the observer frame. However, since our electron
velocity distribution averages to zero, a photon has the same
probability to undergo up or down-scattering. Still, a net
up-scattering (and in turn the formation of a hard tail in the
spectrum) is expected if the magnetic field is inhomogeneous.  For a
photon propagating from high to low magnetic fields, multiple resonant
cyclotron scattering will, on average, up-scatter in energy the
transmitted radiation, while the dispersion in energy decreases with
optical depth (Lyutikov \& Gavriil 2006).  Photon boosting by particle
thermal motion in Thomson limit occurs due to the spatial variation of
the magnetic field and differs qualitatively from the (more familiar)
non-resonant Comptonization (Kompaneets 1956). As a result, the
emerging spectrum is non-thermal and under certain circumstances can
be modeled with two-component spectral models consisting of a
blackbody plus a power-law (Lyutikov
\& Gavriil 2006).


\subsection{The XSPEC implementation of the RCS model}
\label{sec:xspec}

\begin{figure*}
\centering{
\hspace{0.1cm}
\hbox{
\psfig{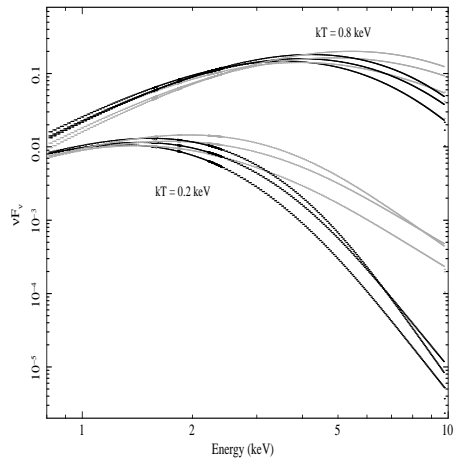}}
\caption{Distorsion of a seed blackbody spectrum through
resonant cyclotron scattering onto magnetosferic electrons, for
two values of the blackbody temperature, 0.2~keV and 0.8~keV. Black
lines show the
RCS model for $\beta_{T}=0.2$ and $\tau_{res} =$ 2, 4, 8 (from bottom to
top), while grey lines are relative to $\beta_{T}=0.4$ and $\tau_{res} =$ 2,
4, 8 (from bottom to top). The normalizations of the various curves are
arbitrary.}}

\label{figrcs}
\end{figure*}

In order to implement the RCS model in {\tt XSPEC}, we created a grid
of spectral models for a set of values of the three parameters
$\beta_{T}$, $\tau_{res}$ and $T$. The parameter ranges are
$0.1\leq\beta_{T}\leq0.5$ (step 0.1; $\beta_T$ is the thermal velocity
in units of $c$), $1
\leq\tau_{res}\leq 10$ (step 1; $tau_{res}$ is the optical depth) 
and $0.1$~keV$\leq T\leq1.3$~keV (step 0.2\,keV; $T$ is the
temperature of the seed thermal surface radiation, assumed to be a
blackbody). For each model, the spectrum was computed in the energy
range 0.01--10\,keV (bin width 0.05\,keV). The final {\tt XSPEC} {\tt
atable} spectral model has therefore three parameters, plus the
normalization constant, which are simultaneously varied during the
spectral fitting following the standard $\chi^2$ minimization
technique. In Fig.\,\ref{figrcs} we show the comparison between a
blackbody model and our RCS model. We stress again that our model has
the same number of free parameters (three plus the normalization) than
the blackbody plus power-law or two blackbody models ($\beta_{T}$,
$\tau_{res}$, $T$, plus the normalization, compared to $kT$, $\Gamma$
(or $kT_2$), plus two normalizations); it has then the same
statistical significance. We perform in the following section a
quantitative comparison between the RCS model and other models
commonly used in the soft X-ray range. However, note that here the RCS
model is meant to model spectra in the 0.1--10\,keV energy range. For
all sources with strong emission above $\sim 20$~keV, the spectrum was
modeled by adding to the RCS a power-law meant to reproduce the hard
tail (see \S\,\ref{results} for details). This power-law does not have
(yet) a clear physical meaning in our treatment, but since it
contributes also to the 0.1--10\,keV band, our RCS parameters depend
on the correct inclusion of this further component.

\section{Observations and Data Analysis}
\label{obs}

Before discussing our data analysis, we would like to outline the
choices we made in selecting the datasets to be used in this work.
Aim of this paper is to show how the RCS model can account for the
X-ray spectra of both steady and variable AXPs and SGRs. Detailed
spectral modeling requires high-quality data and this led us to
consider only the highest signal-to-noise ratio datasets available to
date for these sources. We selected then only those magnetar
candidates having \XMM\, spectra with a number of counts $> 10^5$ and
did not include short ($<10$\,ks) \XMM\, exposures\footnote{Except for
\kes, for which only a single short \XMM\, observation is available.},
\CXO\, or \Swift\, observations. Fortunately most of the magnetars
met the above criterion, but our choice resulted in the exclusion of
\cxo, \axj, \sgrc, \sgrd and \sgre; they are no further considered in
the present investigation\footnote{While this paper was approaching
completion, Tiengo, Esposito \& Mereghetti (2008, ApJ submitted)
reported a detailed 0.1--10\,keV spectrum for \cxo. In their paper,
the successful application of our RCS model to this source is
presented.}.  The remaining sources are divided into three groups, as
follows.

\begin{itemize}
\item{} A set of AXPs which emit in the hard X-ray range, and also happen
to be ``steady'' emitters or showing moderate flux and spectral
variability (flux changes less than a factor of 5; with the exception
of \ea, see also below). These long-term changes are not considered in
the following (see \S~\ref{res:hard} for details). This group
comprises: \uu, \rxs, \kes, and \ea.  When more than one \XMM\,
observation was available, we chose the dataset with the longest
exposure time and least affected by background flares.

\item{} A set of ``transient'' AXPs (often labeled TAXPs), which
includes \xte, \1e, and \wes. To these we add \ee, in the light of the
recent detection of large outbursts from this source (Mereghetti et
al. 2004; Gavriil et al. 2006; Tam et al. 2007; Campana \& Israel
2007), and of its spectral similarities with canonical TAXPs. In order
to follow the spectral evolution without being encumbered with
unnecessary details, we selected only three \XMM\, spectra for each
source, also when more observations were available (e.g. for \ee\, and
\xte). The three chosen datasets correspond to the two most diverse
spectra and to an ``intermediate'' state.

\item{} A set of SGRs, which comprises \sgra (three observations
covering epochs before and after the giant flare of 2004 December 27),
and \sgrb.

\end{itemize}

For all the sources in the first group (except \ea) and for \sgrb\, we
also considered \INT\ data. Although \INT\ and \XMM\ observations were
not always simultaneous, the absence of large spectral variability in
these sources justifies our choice. In particular, for \sgrb\ care has
been taken to select data within periods in which the source was
relatively steady. Although AXP \ea\, and \sgra\, have been also
detected above 20\,keV (Kuiper et al. 2006; Mereghetti et al. 2005b;
Molkov et al. 2005), the \INT\, X-ray counterpart of the former is too
faint to extract a reliable spectrum, while the highly variable hard
and soft X-ray spectrum of the latter, together with the non
simultaneity of the \XMM\, and \INT\, observations, would make any
attempt to model its 1--200\,keV spectral energy distribution
meaningless.

The following subsections provide some details on the observations and
data analysis; a comprehensive log, with the exposure times and epochs
of each observation, is provided in Tab.\,\ref{logobs}.

\subsection{XMM-Newton: soft X-rays}
\label{xmm}

All soft X-ray spectra were collected by the \XMM\, EPIC-pn instrument
(Jansen et al. 2001; Str\"uder et al. 2001), which has the largest
sensitivity in the 1-10~keV band. In order to have a homogeneous
sample of spectra, we re-analysed all the data using the latest SAS
release 7.1.0.  We employed the most updated calibration files
available at the time the reduction was performed (August 2007).
Standard data screening criteria (e.g. cleaning for background flares)
were applied in the extraction of scientific products. We used
FLAG$=0$ and PATTERN between $0-4$ (i.e. single and double events) for
all the spectra. We have checked that spectra generated with only
single events (i.e. PATTERN$=0$) agreed (apart from normalization
factors) with those generated from single and double events.  All the
EPIC-pn spectra were rebinned before fitting, using at least 30 counts
per bin and not oversampling the resolution by more than a factor of 3
(see Rea et al.~2005, 2007c for further details on our \XMM\ data analysis
and reduction).

\subsection{INTEGRAL: hard X-rays}
\label{integral}

In order to take into account in our spectral modeling the
contribution of the hard X-ray emission of \uu, \rxs, \kes and
\sgrb , we used  the hard X-ray spectra derived from \INT\, data.
We selected and analyzed all publicly available IBIS (Ubertini et
al. 2003) pointings, making use of ISGRI  (Lebrun et
al. 2003), the IBIS low energy detector array working in the
15\,keV--1\,MeV energy range. Data were collected for all pointings
within 12$^{\circ}$ from the direction of each source, for a total
2544, 1351, 1894 and 1535 pointings of 2-3\,ks each, for the three AXPs
and the SGR, respectively.  Given the low hard X-ray flux of these
sources, we added all the data in order to have statistically
significant detections.

We processed the data using the Offline Scientific Analysis (OSA)
software provided by the \INT\, Science Data Centre (ISDC) v6.0. We
produced the sky images of each pointing in 10 energy bands between 20
and 300 keV, and added them in order to produce a mosaicked image.
Due to the faintness of the sources we could not derive their spectra
from the individual pointings, so following e.g. G\"otz et al. (2007),
we used the count rates of the mosaicked images to build the time
averaged spectrum of each source.

\section{Spectral Analysis and Results}
\label{results}

All the fits have been performed using {\tt XSPEC} version 11.3 and
12.0, for a consistency check. A 2\% systematic error was added to the
data to partially account for uncertainties in instrumental
calibrations. A constant function has been fitted when using both
\XMM\, and \INT\, data to account for inter-calibration uncertainties
(the values of the constant in the Tables are relative to \XMM\, set
to unity). The 0.5--1\,keV energy range was excluded from our spectral
fitting because: \textit{i)} this is the band where most of the
calibration issues lay (Haberl et al. 2004), and \textit{ii)} emission
in this energy range is mostly affected by interstellar absorption,
and by the choice of the assumed solar abundances. Given the high
column density of all magnetars, and the large uncertainties in the
abundances (probably not even solar) in their directions, this may
lead to spurious features. We checked that for all our targets, the
values of $N_H$ derived fitting the 1--10\,keV EPIC-pn spectra, are
consistent (within the errors) with those obtained using the
0.5--10\,keV range in the same data set. We notice that the absorption
value derived here for the blackbody plus power-law or two blackbodies
models is, on average, slightly higher than that reported in the
literature for the same model.  This is due to our choice of using the
more updated solar abundances by Lodders (2003), instead of the older
ones from Anders \& Grevesse (1989). This does not affect the other
spectral parameters, which are in fact consistent with those
previously published for the same data sets. For all the fits we used
photoelectric cross-sections derived from Balucinska-Church \&
McCammon (1992).

We raise the caveat that no attempt has been made here to distinguish
the pulsed from the non-pulsed emission of these objects, and to model
the spectral variability with phase observed in most of these
sources. This will be the subject of a future investigation.

\subsection{AXPs: the hard X-ray emitters}
\label{res:hard}

In this section we first consider the AXPs with detected hard X-ray
emission, which also coincides with the marginally variable AXPs, with
the exception of \ea\, (Kaspi et al.~2003; Woods et al.~2004; see
below). We recall that, strictly speaking, these hard X-ray emitting
AXPs are not ``steady'' X-ray emitters. Subtle flux and spectral
variability was discovered in \rxs\, and \uu. In particular, \rxs\,
showed a long term, correlated intensity-hardness variability (both in
the soft and hard X-rays), most probably related to its glitching
activity (Rea et al. 2005; Campana et al. 2007; G\"otz et al. 2007;
Dib et al. 2007a; Israel et al.~2007a). \uu\, showed a flux increase
of $\sim 10\%$ (also correlated with a spectral hardening) following
the discovery of its bursting activity (Dib et al.~2007b; Gonzalez et
al. 2007).  Furthermore, thanks to a large
\RXTE\ monitoring campaign, long-term spin period variations and
glitches were discovered in \uu\, \rxs, and \kes, i.e. the three AXPs
which are the brightest both in the soft and hard X-ray bands (Gavriil
\& Kaspi 2002; Dall'Osso et al. 2003; Dib et al. 2007a; Israel et al. 2007a).

Since these flux variations are rather small, we have chosen to model
only the \XMM\, observation closest to the \INT\, one (for
\rxs\ only one \XMM\, observation is available though). Our results
from the spectral modeling of the 1--200\,keV spectrum of
\uu, \rxs, and \kes\, are summarized in Table \ref{tablexmmintegral}
and shown in Fig. \ref{spectraxmmintegral}.

The case of \ea\ is rather different: it showed a large outburst (more
than one order of magnitude flux increase) detected by \RXTE, during
which also bursting activity was detected (Kaspi et
al. 2003). However, in the \XMM\, observations pre and post outburst,
the source showed fluxes which differ only by a factor of 3 (Woods et
al. 2004).  Furthermore, it was observed to emit up to $\sim$30\,keV
by the HEXTE instrument on board of \RXTE\, (Kuiper et al.~2006) and
by \INT, but unfortunately it is too faint in the latter observation
to extract a spectrum.  We then decided to model only the deepest
\XMM\, observation taking into account of the $>$10\,keV component by
adding a power-law with photon index fixed at the HEXTE value (Kuiper
et al.~2006). This is because sizable residuals are present at the
highest energies when the \XMM\ spectrum is modeled either with the
BB+PL or the RCS model. A satisfactory fit requires, in both cases,
the addition of a hard X-ray power-law component (see also Table
\ref{table2259} and Fig. \ref{spectra2259}).

Summarizing, the only source that can be considered (so far) a genuine
``steady'' X-ray emitter, among the AXPs with hard X-ray emission is
\kes. It is interesting to note that this is also the only AXP for which a
blackbody plus a single power-law reproduces well the entire 1-200~keV
spectrum, while for the other hard X-ray emitting AXPs two power-laws are
required. In this respect, the spectral distribution of \kes\
resembles the one of the SGRs (see also \S\ref{discussion}).

In all cases we found that $N_H$, as derived from the RCS model, is
lower than (or consistent with) that inferred from the BB+2PL fit (or
BB+PL in the case of \kes), and consistent with what derived from
fitting the single X-ray edges of \uu, \ea, and \rxs (Durant \& van
Kerkwijk 2006). This is not surprising, since the power-law usually
fitted to magnetar spectra in the soft X-ray range is well known to
cause an overestimate in the column density\footnote{This is because
the absorption model tends to increase the $N_H$ value in response of
the steep rise of the power-law at low energies, which eventually
diverges approaching E=0.} The surface temperature we derived fitting
the RCS model is systematically lower than the corresponding BB
temperature in the BB+2PL or BB+PL models, and is consistent with
being the same ($\sim 0.33$\,keV) in the four sources. On the other
hand the thermal electron velocity and the optical depth are in the
ranges 0.2--0.4 and 1.0--2.1, respectively. Concerning the hard X-ray
power-law, we find that the photon index is, within the errors, the
same when fitting the RCS or the BB+2PL or BB+PL models (note that for
\ea\, it was kept fixed), while the hard PL normalization is larger in
the RCS case with respect to the BB+2PL model. Both the soft and the
hard X-ray fluxes of all these AXPs derived from the RCS fitting are
consistent with those implied by the usual BB+2PL fitting.

\subsection{AXPs: the ``transients''}
\label{res:trans}

``Transient'' AXPs have been discovered only very recently, when an
increase in the X-ray flux by a factor $\sim 100$ over the value
measured a few years before was observed in \xte\, (Ibrahim et
al.~2004; Gotthelf et al.~2004). Later on, new TAXPs have been
observed showing large flux and spectral variations, e.g. \wes\, (Muno
et al.~2007) and \1e\, (Gelfand \& Gaensler~2007; Camilo et al.~2007a;
Halpern et al.~2007).  Very intriguing is the discovery of pulsed
radio emission correlated with the outbursts of \xte\, and \1e\
(Camilo et al. 2006, 2007a), while so far only upper limits have been
set on the radio emission from \wes, \ee\, and other AXPs (Burgay et
al. 2006, 2007; Camilo et al.  2007b).

It is not clear whether AXPs and TAXPs are indeed two distinct groups
of sources.  During the past few years it has became increasingly
evident that flux variations of different magnitudes also occur in
``steady'' AXPs, possibly related to their bursting and glitching
activity (see \S\ref{res:hard}). Furthermore, bursts have been
observed also during the outbursts of the TAXP \xte\, (Woods et
al. 2005) and \wes\ (Muno et al. 2007), the latter also showing a
large glitch (Israel et al.~2007b). However, in this paper we maintain
the distinction between TAXPs and AXPs, partly for historical reasons,
and partly because the two classes may indeed have different spectral
properties, with the TAXPs being characterized by much softer X-ray
spectra, and by the lack, so far, of detection at energies $>10$\,keV.

The results of the TAXPs spectral modeling are summarized in Tables\,
4, 5, 6, 7,
and shown in Figs. \,\ref{spectra1048}, \ref{spectra1810},
\ref{spectra1547}, \ref{spectra1647}.  Also in this case, we chose to model up 
to three spectra representative of the flux and spectral variability
of these sources. Again, $N_H$ derived with the RCS model is lower
than (or consistent with) that inferred from the more common BB+BB
fitting for \xte, \wes, and \1e, and significantly lower in the case
of the BB+PL model applied to \ee\, (and consistent with that derived
by Durant \& van~Kerkwijk 2006). We also found that the RCS model can
easily account for all the spectral and intensity changes in the
TAXPs.  With the exception of \xte, the surface temperature we derive
for all the TAXPs is lower than, or consistent with, that of the
blackbody in the BB+PL or BB+BB model (for the BB+BB model, we refer
to the BB with the lowest temperature). However, considering only the
RCS model, it is evident for \xte, \1e, and
\wes\, that the outburst state has a high surface temperature which cools
down during the decay, while for \ee\, this trend is less clear.
Furthermore, for all the TAXP but \wes, $\beta_{T}$ increases during
the outburst decay. The behavior of $\tau_{res}$ is less homogeneous:
this parameter decreases with decaying flux in \xte and \ee, remains
qconstant in \1e, and shows an increase during the outburst decay in
the case of \wes. Also for these transient sources, the fluxes derived
by the empirical model and the RCS model are consistent.

\subsection{SGRs}
\label{res:sgrs}

Finally, we consider the 1--10\,keV and 1--200\,keV emission of
\sgra\,(see Table\,8 and Fig.\,8) and
\sgrb\,(see Table\,\ref{table1900} and Fig.\,\ref{spectra1900}), respectively.
It has been already noticed that the hard X-ray emission of SGRs is
quite different from that of AXPs (see
\S\ref{res:hard}). In fact, with the exception
of \kes, the spectra of AXPs show a clear turnover between 10 and 20
keV (see Fig. \ref{spectraxmmintegral}) and the fit requires an
additional spectral component. Instead, the hard X-ray emission of
SGRs seems the natural continuation of the non-thermal component which
is dominant in the 1--10 keV energy range. This is why we can use a BB
(or RCS) plus a single power-law in the entire 1--200\,keV range for
\sgrb, while for the hard X-ray emitting AXPs we were forced to add a
second power-law to the BB+PL model.

Similar considerations hold for \sgra, in which case we model the
1--10\,keV emission by adding a power-law component which is intended
to account for the contribution of the hard X-ray emission in the soft
X-ray range. For the latter SGR we modeled three X-ray observations
taken before and after the Giant Flare of 2004 December 27 (Hurley et
al. 2005; Palmer et al. 2005). We found that the $N_H$ value is
consistent within 1 $\sigma$ between the BB+PL and the RCS+PL
models, and the power-law contribution and the photon index vary among
the three spectra in a similar fashion for the two models. Also,
in the RCS+PL model the surface temperature remains constant within
the errors until before the Giant Flare, and then becomes very low
after one year. Besides the temperature, the spectral variability is
accounted for by changes in the parameters describing the
magnetospheric currents, with $\beta_{T}$ and $\tau_{res}$ varying in
the ranges 0.14--0.5 and in the 2.2--4.3, respectively. 

In the \sgrb\, 1--200\,keV spectrum, we found consistent $N_H$ and
spectral index values between the BB+PL and RCS+PL models, and a RCS
surface temperature significantly lower than the corresponding BB
temperature. In all the SGR observations, the derived fluxes are
consistent among the two models.

\section{Discussion}
\label{discussion}

Before discussing our results and the physics we can derive from our
model, we would like to stress once again that the RCS model involves
a number of simplifications (see \S\ref{rcsmodel}). One is the
assumption of a single temperature surface emission. Current-carrying
charges will hit and heat the star surface, generally inhomogeneously
(TLK). In addition, the emission emerging from the surface is likely
to be non-Plankian. While the presence of an atmosphere on top the
crust of a magnetar remains a possibility (see G\"uver et
al. 2007a,b), its properties, are then likely different from those of
a standard (in radiative and hydrostatic equilibrium) atmosphere on,
e.g., a canonical isolated cooling neutron star (see e.g. Ho \& Lai
2003; van Adelsberg \& Lai 2006). The extreme field and (relatively)
low surface temperature ($\lesssim 0.5\ {\rm keV}$) of magnetar
candidates may also be suggestive of a condensed surface, at least if
the chemical composition is mainly Fe (see Turolla, Zane \& Drake
2004). In the light of these considerations, and in the absence of a
detailed model for the surface emission, and for the atmosphere of
strongly magnetized NSs constantly hit by returning currents, we
restricted ourself to a blackbody approximation for the seed thermal
photons.

In spite of these simplifications, we find that the RCS model can
describe the soft X-ray portion of the whole set of magnetar spectra
we have considered, including the TAXPs variability, by using only
three free parameters (plus a normalization factor). This is the same
number of degrees of freedom required by the blackbody plus power law
model, commonly used to fit this energy band.

\subsection{Magnetar magnetospheric properties}

One of the most interesting outcomes of our analysis is the
measure of the magnetospheric properties of magnetars.  In all
sources, steady and variable ones, the value of $\tau_{res}$ is in the
range of $\sim 1$--$6$. This suggests that the entire
class of sources are characterized by similar properties of scattering
electrons, their density and their (thermal) velocity spread. An
optical depth $\tau_0=\tau_{res}/2$ requires a particle density $n_e$
(see eq.~[\ref{tau0}]) which can be easily inferred considering:
\begin{equation} \tau_0 \approx 1.8
\times 10^{-20} n_e r_{sc} \left(\frac{ 1\, {\rm keV}}{ \hbar
\omega_B}\right)  \, , \label{tau00}
\end{equation}
where $r_{sc}$ is the
radius of the scattering sphere
\begin{equation} r_{sc} \approx 8
R_{NS}\left (\frac{B}{B_{crit}}\right )^{1/3} \left (\frac{ 1\, {\rm
keV}}{ \hbar \omega_B} \right )^{1/3} \, , \label{rsc}
\end{equation}
$R_{NS}$ is the neutron star radius and $B_{crit} \approx 4.4 \times
10^{13}$~G is the quantum critical field.  By taking a typical photon
energy of $\sim 1\ {\rm keV}$, $R_{NS} \sim 10^6$~cm and $B\sim 10
B_{crit}$, we get $n_e \approx 1.5\times 10^{13} \tau_{res}\, {\rm
cm}^{-3}$.  This is several orders of magnitude larger than the
Goldreich-Julian density (Goldreich \& Julian 1969) at the same
distance, $n_{GJ} \approx n_e \pi r_{sc} /(3
\tau_{res} R_{lc}) \sim 2 \times 10^{10}$~cm$^{-3}$ (where $R_{lc}$ is the light
cylinder radius and we took $P\sim 10$~s). While the charge density is
large when compared with the minimal Goldreich-Julian density, it
provides a negligible optical depth to non-resonant Thomson
scattering. Only the resonant cyclotron scattering makes an efficient
photon boosting possible.

Our present model does not include a proper treatment of
magnetospheric currents, so that $\tau_{res}$ is a free parameter
related to the electron density. Nevertheless, it is useful to compare
the values of the optical depth inferred here to those expected when a
current flow arises because a steady twist is implanted in the star
magnetosphere, as in the case investigated by TLK under the assumption
of axysimmetry and self-similarity. If the scattering particles
have a collective motion (bulk velocity $\beta_{bulk}$), the
efficiency of the scattering process is related to $\tau_{res}
\beta_{bulk}$ (e.g. Nobili, Turolla \& Zampieri 1993). This quantity
is shown as a function of the magnetic colatitude in Fig.~5 of TLK for
different values of the twist angle, $\Delta\phi_{N-S}$. By assuming
$\beta_{bulk}=1$ and integrating over the angle, we get the average
value of the scattering depth as a function of $\Delta\phi_{N-S}$,
which is shown in Fig.~\ref{taures}. The curves corresponding to a
different value of $\beta_{bulk}$ can be obtained simply by reading
the quantity shown in Fig.~\ref{taures} as $\tau_{res} \beta_{bulk}$
and by rescaling the $y$-axis. As we can see, a value of $\tau_{res}
\sim 1$ is only compatible with very large values of the twist angle
(i.e $\Delta\phi_{N-S} > 3$), while typical values of $\tau_{res} \sim
2$, as those obtained from some of our fits, require $\beta_{bulk}
\lesssim 0.5$ to be compatible with $\Delta\phi_{N-S}\sim 3$ (the
smaller is $\beta_{bulk}$, the smaller is the value of the twist
angle). This is consistent with the fact that the RCS model has been
computed under the assumption of vanishing bulk velocity for the
magnetospheric currents, and it is compatible with TLK model only when
in the latter it is $\beta_{bulk}\ll 1$.

\subsection{Comparison between AXPs and SGRs}

In the last few years the detection of bursts from AXPs (Gavriil et
al. 2002; Kaspi et al. 2003) strengthened their connection with
SGRs. However, the latter behave differently in many respects.  Below
$\sim4$~keV, the SGRs emission can be described either by a blackbody
or an RCS component. At higher energies though ($>4$\,keV), their
spectra require the addition of a power-law component, which well
describes the spectrum until $\sim 200$\,keV. The non-thermal
component dominates their spectra to the point that the choice of a
blackbody or the RCS model at lower energies does not affect
significantly the value of the hard X-ray power-law index, nor the
energy at which this component starts to dominate the spectrum (see
e.g Tab.\,9 and Fig.\,9). The spectra of SGRs are then strongly
non-thermally dominated in the 4--200\,keV range.

The case of the AXPs is different (with the exception of
\kes, see below). These sources show a more complex spectrum, with an
evident non-thermal component below $\sim10$~keV, apparently different
from that observed at higher energies. For the AXPs detected at
energies $>$20\,keV, the spectrum can be described by a RCS component
until 5--8\,keV, above which the non-thermal hard X-ray component
becomes important, and (e.g. for \rxs\, and \uu) dominates until $\sim
200$\,keV. In the case of the BB+2PL model instead, the non-thermal
component responsible for the hard X-ray part of the spectrum starts
to dominate only above $\sim 10$~keV (see e.g.
Figs. \ref{spectraxmmintegral} and \ref{spectra2259}).  This is
important, because the measurement of a down-break of the hard X-ray
power-law has remarkable physical implications and may prove useful in
constraining the physical parameters of the model for the hard X-ray
emission.  It is worth noting that the photon index of the hard X-ray
component in AXPs does not strongly depend on the modeling of the
spectrum below 10\,keV, while, its normalization and, as a
consequence, the value at which the hard tail starts to dominate the
spectrum, do.

In this picture \kes\, seems an exception. From the spectral point of
view, \kes\ appears as the more SGR-like among the AXPs. Its
multi-band spectrum can be well fitted by a BB+PL or RCS+PL model,
with parameters very similar to SGRs (compare Figs.~\ref{spectra1900},
\ref{spectraxmmintegral} and Tables~\ref{spectra1900},
\ref{spectraxmmintegral}). This may suggest that this
source is a potential transition object between the two
classes. However, at variance with the SGRs, this source seems to be
the least active bursters among AXPs. Note that, at variance with the
other magnetars, in the case of the two SGRs and \kes, our model
requires two additional free parameters, with respect to the BB+PL, to
account for the hard X-ray power-law.

The fact that hard X-ray spectra detected from AXPs are much flatter
than those of SGRs may also suggest a possible difference in the
physical mechanism that powers the hard tail in the two classes of
sources. Within the magnetar scenario, Thompson \& Beloborodov (2005)
discussed how soft $\gamma$-rays may be produced in a twisted
magnetosphere, proposing two different pictures: either thermal
bremsstrahlung emission from the surface region heated by returning
currents, or synchrotron emission from pairs created higher up ($\sim$
100 km) in the magnetosphere. Moreover, a third scenario involving
resonant magnetic Compton up-scattering of soft X-ray photons by a
non-thermal population of highly relativistic electrons has been
proposed by Baring \& Harding (2007). It is interesting to note that
3D Monte Carlo simulations (Fernandez \& Thompson 2007; Nobili,
Turolla \& Zane 2008) show that multiple peaks may appear in the
spectrum. In particular, in the model by Nobili, Turolla \& Zane
(2008), a second ``hump'' may be present when up-scattering is so
efficient that photons start to fill the Wien peak at the typical
energy of the scattering electrons. The change in the spectral slope
may be due, in this scenario, to the peculiar, ``double-humped'' shape
of the continuum. The precise localization of the down-break is
therefore of great potential importance and might provide useful
information on the underlying physical mechanism responsible for the
hard emission.

The RCS model applied to the evolution of the outbursts of the TAXPs
known up to now shows how the outburst may results from a heating of
the NS surface, which slowly cools in a timescale of
months/years. AXPs outbursts are thought to be caused by large scale
rearrangement of the surface/magnetospheric field, either accompanied
or triggered by fracturing of the NS crust. It is worth noticing that
from our modeling we find that the surface temperature cools down
during the outburst decay, while the magnetospheric characteristics
change in a different way from source to source.


\subsection{Correlations}

The quite large number of observations we analyzed (both relative
to different sources and to single sources in different emission
states) allows to search for possible correlations among the various
quantities, both in the entire sample, i.e. looking at the population
of magnetar candidates at large, and in the time evolution of a single
source.

Fig.\,11 summarizes the results of our spectral fits. The various
panels show how the three model parameters ($T$, $\tau_{res}$ and
$\beta_T$) are related to the X-ray luminosity in the 1--10\,keV band
($L_{\rm 1-10\, keV}$) and to the magnetic field $B$. The latter is
derived from $P$ and $\dot P$, assuming that the magnetic field is a
core-centered dipole and the spin-down is due magnetic dipole
radiation.

An inspection of the panels in Fig.\,11 does not reveal any obvious
correlation for the entire set of observations.  To verify this, we
have run a Spearman rank test and we only found a positive correlation
between $B$ and both $\tau_{res}$ and $\beta_T$ (deviation from the
null hypothesis at about the $93\%$ and $89 \%$ confidence level,
respectively). No correlations with a significance level above $\sim
65\%$ were found in all the other cases. Both parameters $\beta_T$ and
$\tau_{res}$ control the scattering efficiency, but the meaning of
their correlation with the field strength, which seems to be direct in
the case of the optical depth and inverse in the case of the thermal
velocity (Fig.\,11) is not of immediate interpretation. The optical
depth scales as $n_e r/B$ (see eq.~[3]). If we make again a
comparison with the twisted magnetosphere model (TLK), in which $n_e
\propto B/r$, this is not expected. Taken face value, an increase of
the optical depth with increasing $B$ implies that the product $n_er$
grows more rapidly than $B$. Since both in the RCS model and in TLK
the scattering radius is $\propto B^{1/3}$, this implies that $n_e$
should grow faster than what expected in a self-similar magnetostatic
configuration. Furthermore, it is interesting to note

On the other, we caveat that these considerations are
largely model dependent and, in order to assess this issue, a detailed
treatment of the magnetosphere, including more realistic profiles for
the electron density and velocity distribution, is needed.

As discussed earlier, a more interesting trend is found restricting to
observations of the same source at different epochs. In many transient
AXPs (e.g. \xte, \1e, and \wes) we observe a clear correlation between
the surface temperature and the X-ray luminosity, which is expected
since in the RCS model an enhanced surface thermal emission produces
more seeds for resonant up-scattering. However, once again there is no
clear trend relating changes in $\tau_{res}$ and $\beta_T$ to changes
in luminosity for the entire TAXP sample.  In most transient sources
at least one of these two parameters increases with flux, and this may
be enough to guarantee that the spectrum hardens at larger
luminosities, but in no case there is a simultaneous increase or
decrease of both $\tau_{res}$ and $\beta_T$ during the outburst
decay. Whether this is due to a degeneracy in the model parameter
space or it reflects a real trend is not clear at present.

\section{Conclusion}

In this paper we showed that the soft X-ray emission of magnetars can
be explained by resonant cyclotron scattering of their thermal surface
emission by a cloud of hot magnetospheric electrons. This model
satisfactorily reproduces the spectral shape of all magnetars soft
X-ray emission, using the same number of free parameters than the
widely used blackbody plus power-law model (except for the SGRs where
the much harder spectrum below 10\,keV, still requires the addition of
a power-law on top of the resonant cyclotron scattering model, being
the same power-law component responsible for their hard X-ray
emission). This means that the RCS model not only catches the main
features of the thermal and non-thermal components observed in these
sources below $\sim 10$~keV, but also successfully provides a
quantitative interpretation. For the magnetars presenting an hard
X-ray emission we included this further component in order to take
into account in our modeling of the contribution of this component
down to the soft X-ray part of the spectrum.

This work represents one of the first attempts to infer some physical
values from the $1-10$~keV spectra of magnetars. Future refinements
are in progress, in order to improve the RCS model from a 1D
analytical model toward a 3D Monte Carlo based code (as the more
advanced codes developed by Fernandez \& Thompson 2007 and Nobili,
Turolla \& Zane 2008). Furthermore, this model eventually applied to
the detailed spectra that {\it XEUS} and/or {\it Con--X} will possibly
make available in the near future, appear a promising step toward the
complete understanding of the physics behind magnetars soft X-ray
emission.

\vspace{1cm}

We acknowledge Valentina Bianchin and Gavin Ramsay for their help in
building the {\tt XSPEC} RCS model, and Fotis Gavriil for kindly
allowing us to look into his preliminary model. Furthermore, we thank
Gianluca Israel and Andrea Tiengo for useful discussions and key
comments on the preliminary draft. NR is supported by an NWO Veni
Fellowship, and acknowledges the warm hospitality of the Mullard Space
Science Laboratory, where this work was started, and of the Purdue
University where it has been completed. SZ acknowledges STFC for
support through an Advanced Fellowship. D.G. acknowledges the French
Space Agency (CNES) for financial support. This paper is based on
observations obtained with \XMM\ and \INT, which are both ESA science
missions with instruments and contributions directly funded by ESA
Member States and the USA (through NASA). The RCS model is available
to the community on the {\tt XSPEC}
website\footnote{http://heasarc.gsfc.nasa.gov/docs/xanadu/xspec/models/rcs.html}.


\newpage

\begin{table*}[t]
\setlength{\tabcolsep}{0.02in}
\centering
\caption{Log Of The \XMM\, and \INT\, Observations Analysed In This Paper.}
\begin{tabular}{ccc}
\hline
\hline
\multicolumn{3}{c}{\XMM} \\
\hline
 Source & Date (YYYY/MM/DD) & Exposure (ks) \\
\hline
\uu & 2004/03/01& 44 \\
\rxs & 2003/08/28 & 45 \\
\kes & 2002/10/07 & 6 \\
\ea & 2002/06/11 & 52 \\
\ee & 2003/06/16 & 69 \\
    & 2005/06/17 & 32 \\
   & 2007/06/14 &  48 \\
\xte & 2004/09/18 & 28 \\
 & 2005/09/20 &  42 \\
& 2006/03/13 & 51\\
\1e & 2006/08/21 & 47 \\
& 2007/08/09 & 16 \\
\wes  & 2006/09/16 &  80\\
& 2006/09/22 & 20 \\
\sgra & 2003/04/03 &  55 \\
      & 2004/10/06 & 19 \\
& 2005/10/04 & 33 \\
\sgrb & 2005/09/17 &  30  \\
\hline
\multicolumn{3}{c}{\INT} \\
\hline
 Source & Date (YYYY/MM/DD) & Exposure (Ms) \\
\hline
\uu & 2003/03/03-2006/08/13 & 1.9 \\
\rxs & 2003/02/28-2005/10/02 & 2.7 \\
 \kes & 2003/03/10-2006/04/28 & 4.0 \\
\sgrb & 2003/03/06-2006/09/26 & 3.7 \\
\hline
\hline
\end{tabular}
\label{logobs}
\end{table*}



\begin{table*}
\setlength{\tabcolsep}{0.01in}
\centering
\caption{Spectral Parameters: \uu, \rxs, and \kes}
\begin{tabular}{lcccccc}
\hline
\hline
 {\em AXPs} & \multicolumn{2}{c}{4U\,0142+614$^*$} &
 \multicolumn{2}{c}{1RXS\,J1708--4009$^*$} &
 \multicolumn{2}{c}{1E\,1841--045} \\

\hline
 \multicolumn{1}{l}{Parameters}
& BB+2PL & RCS+PL
& BB+2PL & RCS+PL
& BB+PL & RCS+PL   \\

\hline
  N$_{H}$
& 1.67$^{+0.02}_{-0.02}$ & 0.81$^{+0.05}_{-0.05}$
&  1.91$^{+0.06}_{-0.06}$ &  1.67$^{+0.05}_{-0.05}$
& 2.38$^{+0.4}_{-0.1}$ & 2.57$^{+0.13}_{-0.15}$  \\

constant
& 1.01 & 1.10
& 1.05 & 0.80
& 1.02  & 1.09 \\
 & & & & & &  \\

kT (keV)
& 0.43$^{+0.03}_{-0.03}$ & 0.30$^{+0.05}_{-0.05}$
& 0.47$^{+0.01}_{-0.01}$ & 0.32$^{+0.05}_{-0.05}$
& 0.51$^{+0.03}_{-0.02}$ &  0.39$^{+0.05}_{-0.05}$  \\

BB~norm
&  8.7$^{+0.4}_{-0.5}\times10^{-4}$ &
&  2.4$^{+0.1}_{-0.2}\times10^{-4}$ &
& 2.4$^{+0.6}_{-0.3}\times10^{-4}$ & \\

$\Gamma_1$
& 4.14$^{+0.04}_{-0.04}$ &
& 2.70$^{+0.08}_{-0.08}$ &
&  \\

PL$_1$~norm
& 0.30$^{+0.08}_{-0.08}$ &
&  0.016$^{+0.003}_{-0.004}$ &
&  &  \\
 & & & & & &  \\

$\beta_{T}$
&  & 0.33$^{+0.05}_{-0.05}$
& & 0.38$^{+0.03}_{-0.03}$
& & 0.23$^{+0.05}_{-0.05}$  \\

$\tau_{res}$ &  & 1.9$^{+0.2}_{-0.2}$
& & 2.1$^{+0.2}_{-0.2}$
& & 1.13$^{+0.3}_{-0.2}$  \\

RCS~norm
& &  4.5$^{+0.6}_{-0.8}\times10^{-3}$
&  &  8.1$^{+1.1}_{-1.3}\times10^{-4}$
& & 3.1$^{+2.3}_{-1.1}\times10^{-4}$ \\
 & & & & & &  \\

$\Gamma_2$
& 0.78$^{+0.1}_{-0.07}$ &  1.1$^{+0.1}_{-0.1}$
& 0.76$^{+0.1}_{-0.1}$ &  1.0$^{+0.1}_{-0.1}$
& 1.47$^{+0.04}_{-0.05}$ & 1.47$^{+0.05}_{-0.05}$ \\

PL$_2$~norm
& 1.4$^{+0.1}_{-0.1}\times10^{-4}$ & 5.0$^{+0.1}_{-0.1}\times10^{-4}$
& 8.6$^{+0.1}_{-0.1}\times10^{-5}$ & 4.2$^{+0.1}_{-0.1}\times10^{-4}$
& 2.4$^{+0.6}_{-0.6}\times10^{-3}$& 2.2$^{+0.1}_{-0.1}\times10^{-3}$\\
 & & & & & & \\

Flux 1--10\,keV
& 1.1$^{+0.8}_{-0.8}\times10^{-10}$ & 1.1$^{+0.8}_{-0.8}\times10^{-10}$
&2.6$^{+0.3}_{-0.3}\times10^{-11}$ & 2.6$^{+1.1}_{-0.8}\times10^{-11}$
& 2.2$^{+0.2}_{-0.3}\times10^{-11}$ & 2.1$^{+0.2}_{-0.3}\times10^{-11}$ \\

Flux 1--200\,keV
&2.3$^{+1.7}_{-1.1}\times10^{-10}$ & 2.3$^{+1.0}_{-1.3}\times10^{-10}$
&  1.1$^{+0.5}_{-0.5}\times10^{-10}$ & 1.4$^{+0.8}_{-0.8}\times10^{-10}$
& 1.1$^{+0.8}_{-0.8}\times10^{-10}$ & 1.1$^{+0.8}_{-0.6}\times10^{-10}$  \\
 & & & & & & \\

$\chi^2_{\nu}$ (dof)
&  0.99 (216) & 0.80 (216)
& 1.11 (202) & 1.01 (202)
& 1.14 (158) & 1.08 (156) \\

\hline
\hline
\end{tabular}
\tablecomments{Best fit values of the spectral parameters
obtained by fitting the $\sim$1--200\,keV \XMM\, and \INT\, AXPs'
spectra with a blackbody plus two power-laws model (BB+2PL) for
\uu\, and \rxs, while a single power-law was used for \kes. Furthermore, all the
sources were modeled with a resonant cyclotron scattering model plus a
power-law (RCS+PL). Errors are at 1$\sigma$ confidence level, reported
fluxes are absorbed and in units of \ergscm2 , and N$_{H}$ in units of
$10^{22}$\,cm$^{-2}$ and assuming solar abundances from Lodders
(2003); 2\% systematic error has been included. See also
Fig.\,\ref{spectraxmmintegral} and
\S\,\ref{res:hard} for details. $^{*}$: source slightly variable in
flux and spectrum, see text for details.}
\label{tablexmmintegral}
\end{table*}


\clearpage


\begin{figure*}
\centering{
\hspace{0.1cm}
\vbox{
\hbox{
\psfig{figure=f2a.eps,height=3.5cm,angle=270,width=7cm}
\psfig{figure=f2b.eps,height=3.5cm,angle=270,width=7cm}}
\hbox{
\psfig{figure=f2c.eps,height=3.5cm,angle=270,width=7cm}
\psfig{figure=f2d.eps,height=3.5cm,angle=270,width=7cm}}
\hbox{
\psfig{figure=f2e.eps,height=3.5cm,angle=270,width=7cm}
\psfig{figure=f2f.eps,height=3.5cm,angle=270,width=7cm}}
\hbox{
\psfig{figure=f2g.eps,height=3.5cm,angle=270,width=7cm}
\psfig{figure=f2h.eps,height=3.5cm,angle=270,width=7cm}}
\hbox{
\psfig{figure=f2i.eps,height=3.5cm,angle=270,width=7cm}
\psfig{figure=f2l.eps,height=3.5cm,angle=270,width=7cm}}
\hbox{
\psfig{figure=f2m.eps,height=3.5cm,angle=270,width=7cm}
\psfig{figure=f2n.eps,height=3.5cm,angle=270,width=7cm}} } }
\caption{\uu, \rxs\, and \kes: left column shows the spectra in Counts/s/keV
while in the right column we report the $\nu$F$_\nu$ plots. For \uu\,
and \rxs\, the upper panels are relative to the modeling with a
blackbody plus two power-laws (BB+2PL), while we used a blackbody plus
power-law for \kes. Bottom panels report for all the sources the
resonant cyclotron scattering plus a power-law model (RCS+PL). See
Tab.\,\ref{tablexmmintegral} and \S\,\ref{res:hard} for details. 
\label{spectraxmmintegral}}
\end{figure*}


\newpage


\begin{figure*}[t]
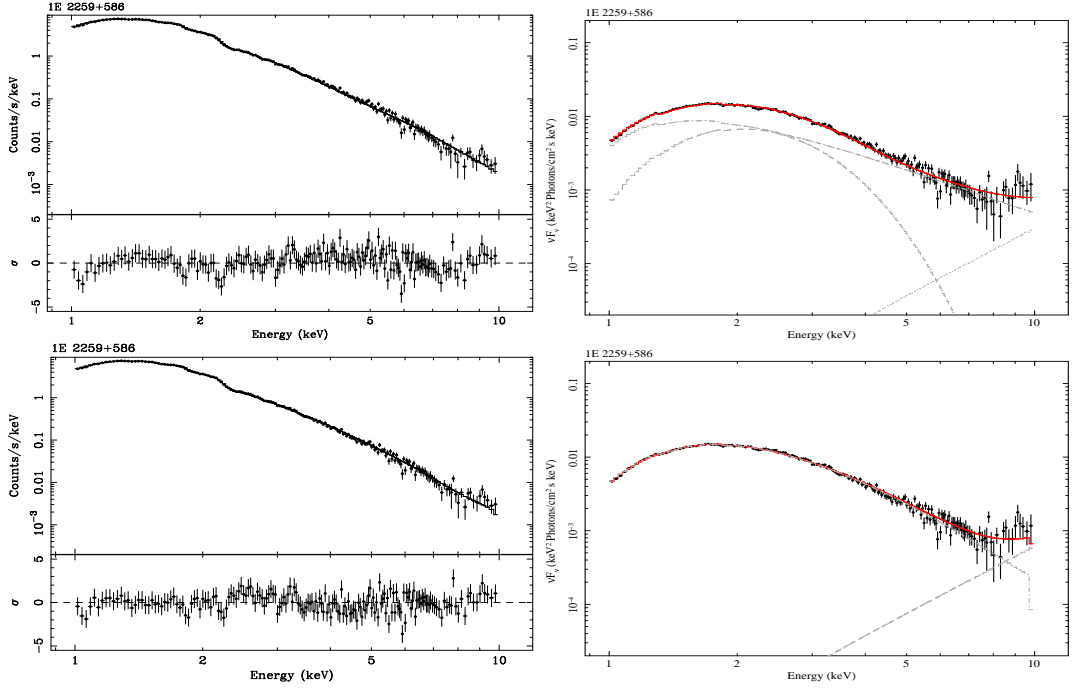

\centering{
\hspace{0.1cm}
\vbox{
\hbox{
\psfig{figure=f3a.eps,height=4.5cm,width=7cm,angle=270}
\psfig{figure=f3b.eps,height=4.5cm,width=7cm,angle=270}}
\hbox{
\psfig{figure=f3c.eps,height=4.5cm,width=7cm,angle=270}
\psfig{figure=f3d.eps,height=4.5cm,width=7cm,angle=270}}}}
\caption{\ea: left column shows the spectra in Counts/s/keV while in the right column
we report the $\nu$F$_\nu$ plots. The upper panels are relative to the
modeling with a blackbody plus two power-laws (BB+2PL), while bottom
panels report the resonant cyclotron scattering plus a power-law model
(RCS+PL). Note the hard X-ray spectrum has been fixed at the value
from Kuiper et al.~(2006). See Tab.\,\ref{table2259} and
\S\,\ref{res:hard} for details.
\label{spectra2259}}
\end{figure*}


\begin{table*}
\setlength{\tabcolsep}{0.01in}
\centering
\caption{Spectral Parameters: \ea}
\begin{tabular}{lcc}
\hline
\hline
  {\em AXP} & \multicolumn{2}{c}{1E\,2259+586$^{*}$} \\
\hline
\multicolumn{1}{l}{Parameters}
& BB+2PL & RCS+PL \\
\hline

N$_{H}$
& 0.97$^{+0.04}_{-0.03}$ & 0.89$^{+0.02}_{-0.02}$ \\
 & &    \\

kT (keV)
& 0.41$^{+0.03}_{-0.03}$ & 0.32$^{+0.02}_{-0.02}$ \\

BB~norm
& 2.77$^{+0.02}_{-0.01}\times10^{-4}$ &  \\

$\Gamma_1$
& 3.98$^{+0.03}_{-0.02}$  &  \\

PL$_1$~norm
& 4.89$^{+0.04}_{-0.04}\times10^{-2}$ & \\
 & &  \\

$\beta_{T}$
& & 0.32$^{+0.03}_{-0.03}$ \\

$\tau_{res}$
& &  1.0$^{+0.2}_{-0.2}$ \\

RCS~norm
& &  1.0$^{+0.1}_{-0.1}\times10^{-3}$ \\
& &  \\

$\Gamma_2$
&  1.02 &  1.02  \\

PL$_2$~norm
&  1.65$^{+1.0}_{-1.0}\times10^{-7}$ &  5.0$^{+1.0}_{-1.0}\times10^{-5}$ \\
& &  \\

Flux 1--10\,keV
& 2.5$^{+0.1}_{-0.1}\times10^{-11}$ & 2.5$^{+0.1}_{-0.1}\times10^{-11}$  \\
 & &  \\

$\chi^2_{\nu}$ (dof)
&  1.15 (178) &   0.94 (178) \\

\hline
\hline
\end{tabular}
\tablecomments{Best fit values of the spectral parameters obtained
by fitting the $\sim$1--10\,keV \XMM\, observation of \ea\, with a
blackbody plus two power-laws model (BB+2PL), and with a resonant
cyclotron scattering model plus a power-law (RCS+PL). We fixed the
second power-law photon index to $\Gamma_2=1.02$, the value reported
in Kuiper et al.~(2006) from \RXTE\, measurements.  Errors are at
1$\sigma$ confidence level, reported fluxes are absorbed and in units
of \ergscm2 , and N$_{H}$ in units of $10^{22}$\,cm$^{-2}$ and
assuming solar abundances from Lodders (2003); 2\% systematic error
has been included. See also Fig.\,\ref{spectra2259} and
\S\,\ref{res:hard} for details. $^{*}$: source variable in flux and
spectrum, see text for details.}
\label{table2259}
\end{table*}


\newpage


\begin{figure*}
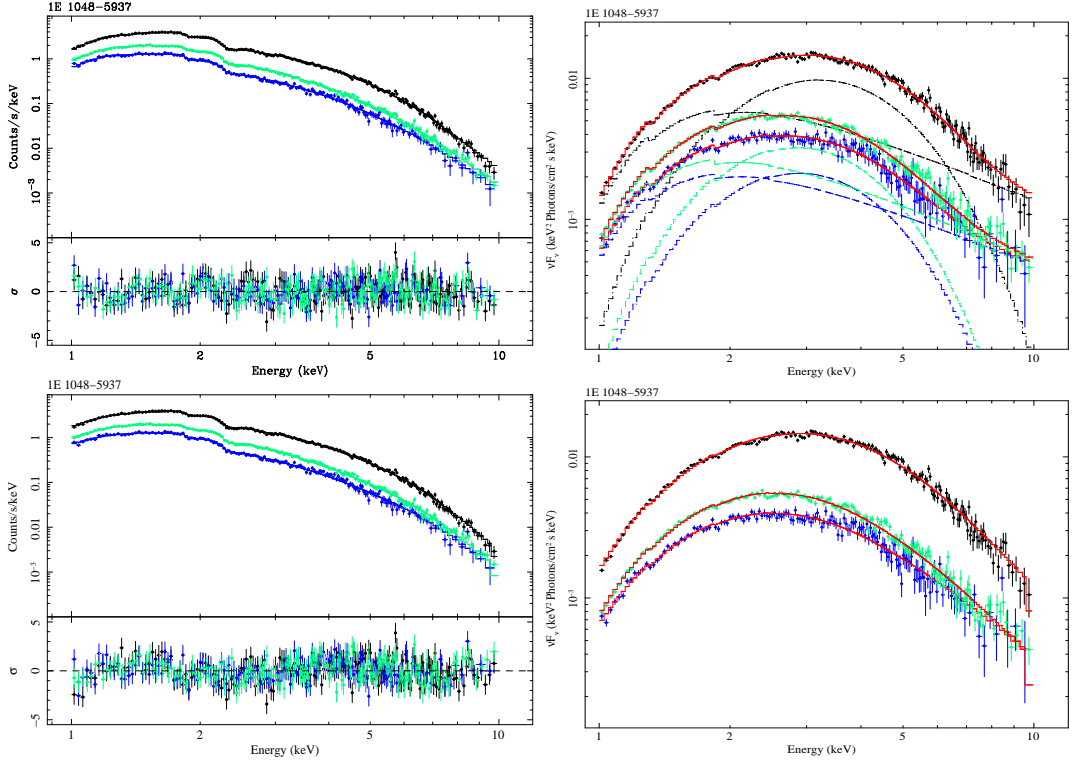

\centering{
\hspace{0.1cm}
\vbox{
\hbox{
\psfig{figure=f4a.eps,height=5cm,angle=270,width=7cm}
\psfig{figure=f4b.eps,height=5cm,angle=270,width=7cm}}
\hbox{
\psfig{figure=f4c.eps,height=5cm,angle=270,width=7cm}
\psfig{figure=f4d.eps,height=5cm,angle=270,width=7cm}}}}
\caption{\ee: left column represents the spectra in Counts/s/keV while in the
right column we report the $\nu$F$_\nu$ plots. The upper panels are
relative to the modeling with a blackbody plus one power-law (BB+PL),
while bottom panels report the resonant cyclotron scattering model
(RCS). See Tab.\,\ref{table1048} and \S\,\ref{res:trans} for
details. Black, blue, and light-green colors are relative to
observations taken in 2007, 2005 and 2003, respectively. The red lines
represent the total model, while the dashed lines are the single
components.
\label{spectra1048}}
\end{figure*}


\begin{table*}
\setlength{\tabcolsep}{0.01in}
\centering
\caption{Spectral Parameters: \ee}
\begin{tabular}{lcccccc}
\hline
\hline
  {\em AXP} & \multicolumn{6}{c}{\ee} \\
\hline
 & \multicolumn{2}{c}{2003}
& \multicolumn{2}{c}{2005}
&  \multicolumn{2}{c}{2007} \\
\hline

\multicolumn{1}{l}{Parameters}
& BB+PL & RCS
& BB+PL & RCS
& BB+PL  & RCS \\
\hline

N$_{H}$
& 1.68$^{+0.03}_{-0.03}$ & 0.98$^{+0.04}_{-0.04}$
& 1.56$^{+0.05}_{-0.04}$& 0.73$^{+0.04}_{-0.04}$
& 1.71$^{+0.04}_{-0.03}$ & 0.82$^{+0.05}_{-0.05}$\\
 & & & & & &   \\

kT (keV)
& 0.63$^{+0.02}_{-0.02}$ & 0.39$^{+0.04}_{-0.04}$
& 0.64$^{+0.03}_{-0.04}$ & 0.44$^{+0.05}_{-0.04}$
& 0.73$^{+0.01}_{-0.01}$ & 0.45$^{+0.05}_{-0.05}$\\

BB~norm
& 1.01$^{+0.05}_{-0.05}\times10^{-4}$ &
&  0.7$^{+0.1}_{-0.1}\times10^{-4}$   &
& 3.00$^{+0.08}_{-0.08}\times10^{-4}$ & \\

$\Gamma_1$
& 3.31$^{+0.02}_{-0.04}$ &
& 3.18$^{+0.03}_{-0.04}$ &
& 3.20$^{+0.07}_{-0.07}$ & \\

PL$_1$~norm
&  1.10$^{+0.13}_{-0.04}\times10^{-2}$ &
& 0.7$^{+0.1}_{-0.2}\times10^{-2}$&
& 2.2$^{+0.1}_{-0.1}\times10^{-2}$ & \\
& & & & & & \\

$\beta_{T}$
& & 0.29$^{+0.02}_{-0.02}$
& & 0.35$^{+0.02}_{-0.04}$
&  & 0.29$^{+0.05}_{-0.05}$ \\

$\tau_{res}$
& &  2.7$^{+0.2}_{-0.4}$
& & 2.0$^{+0.1}_{-0.5}$
& & 4.7$^{+0.2}_{-0.2}$ \\

RCS~norm
& &  1.9$^{+0.1}_{-0.1}\times10^{-4}$
& & 1.01$^{+0.08}_{-0.11}\times10^{-4}$
& & 3.0$^{+0.1}_{-0.1}\times10^{-4}$ \\
 & & & & & &  \\

Flux 1--10\,keV
& 1.1$^{+0.4}_{-0.4}\times10^{-11}$& 1.1$^{+0.4}_{-0.4}\times10^{-11}$
& 0.8$^{+0.3}_{-0.4}\times10^{-11}$ & 0.8$^{+0.4}_{-0.4}\times10^{-11}$
& 3.0$^{+0.5}_{-0.4}\times10^{-11}$ & 3.0$^{+0.7}_{-0.6}\times10^{-11}$ \\
 & & & & & &  \\

$\chi^2_{\nu}$ (dof)
& 0.99 (176) & 0.98 (176)
&  0.99 (153)  &   1.00 (153)
& 1.08 (184) & 1.23 (184)\\

\hline
\hline
\end{tabular}
\tablecomments{Best fit values of the spectral parameters obtained
by fitting several $\sim$1--10\,keV \XMM\, spectra, taken in different
source states, with a blackbody plus power-law model (BB+PL), and with
a resonant cyclotron scattering model (RCS). Errors are at 1$\sigma$
confidence level, reported fluxes are absorbed and in units of
\ergscm2 , and N$_{H}$ in units of $10^{22}$\,cm$^{-2}$ and assuming
solar abundances from Lodders (2003); 2\% systematic error has been
included. See also Fig.\,\ref{spectra1048} and \S\,\ref{res:trans} for
details.}
\label{table1048}
\end{table*}



\begin{figure*}
\hspace{0.1cm}
\centering{
\vbox{
\hbox{
\psfig{figure=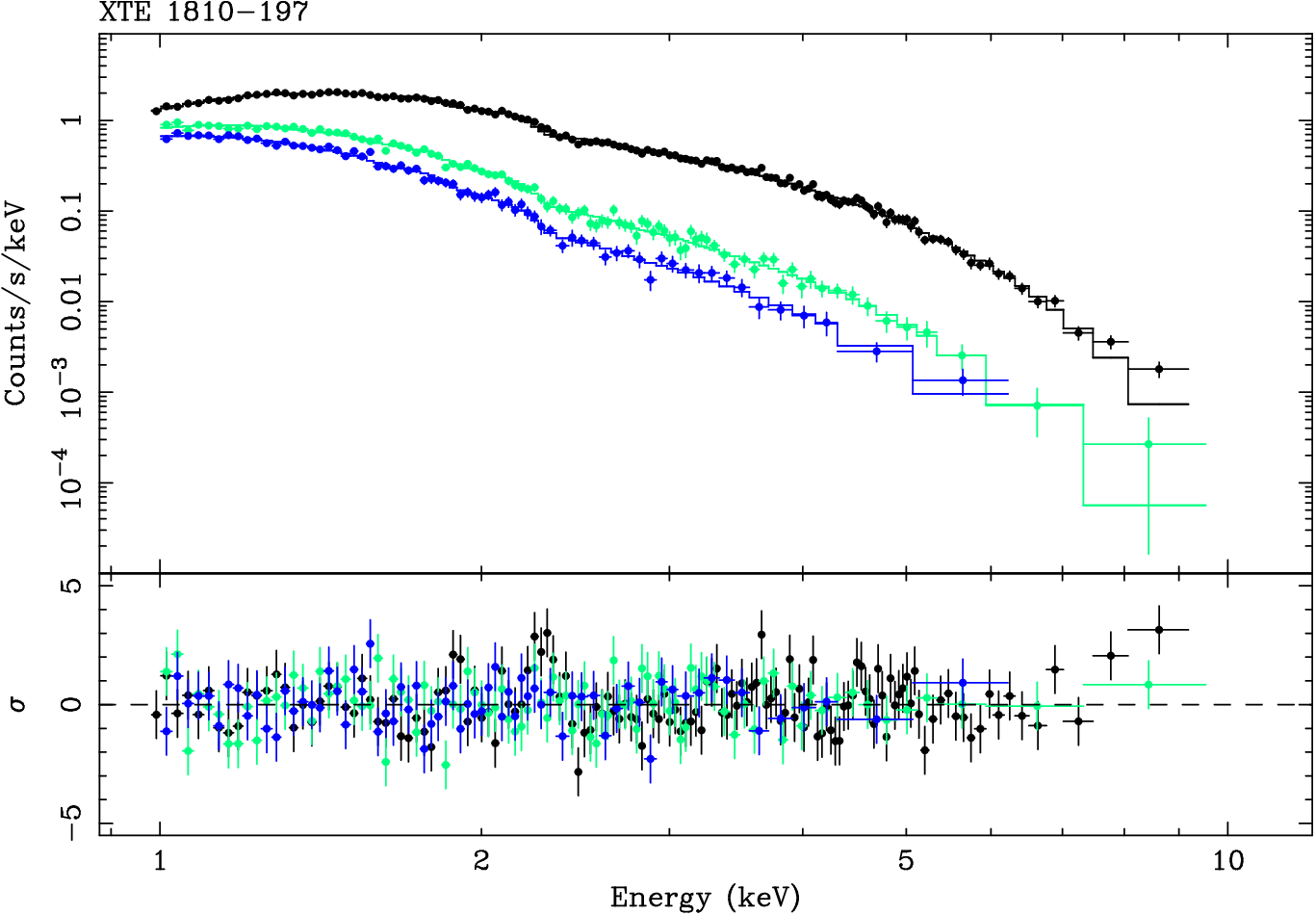,height=5cm,angle=270,width=7cm}
\psfig{figure=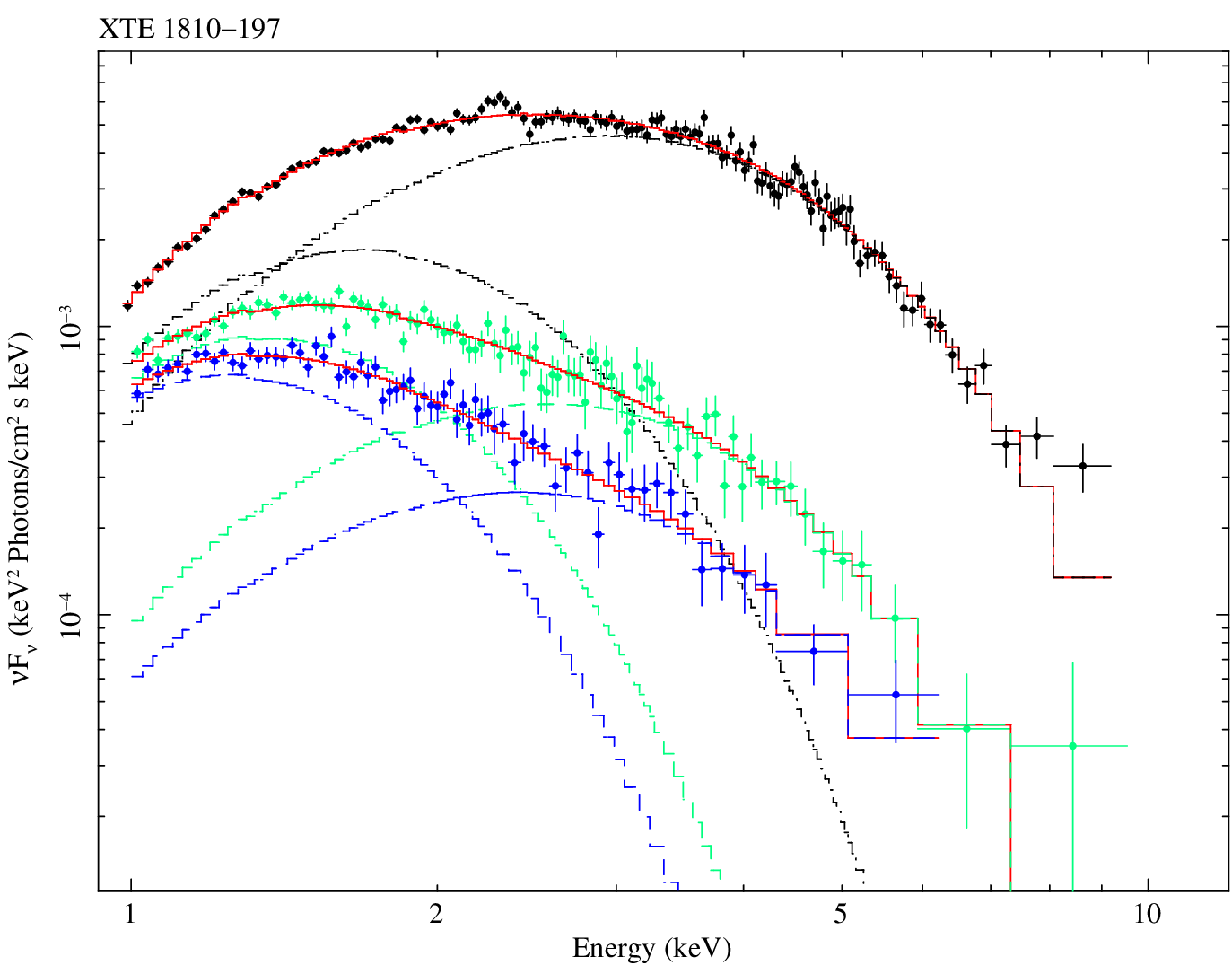,height=5cm,angle=270,width=7cm}}
\hbox{
\psfig{figure=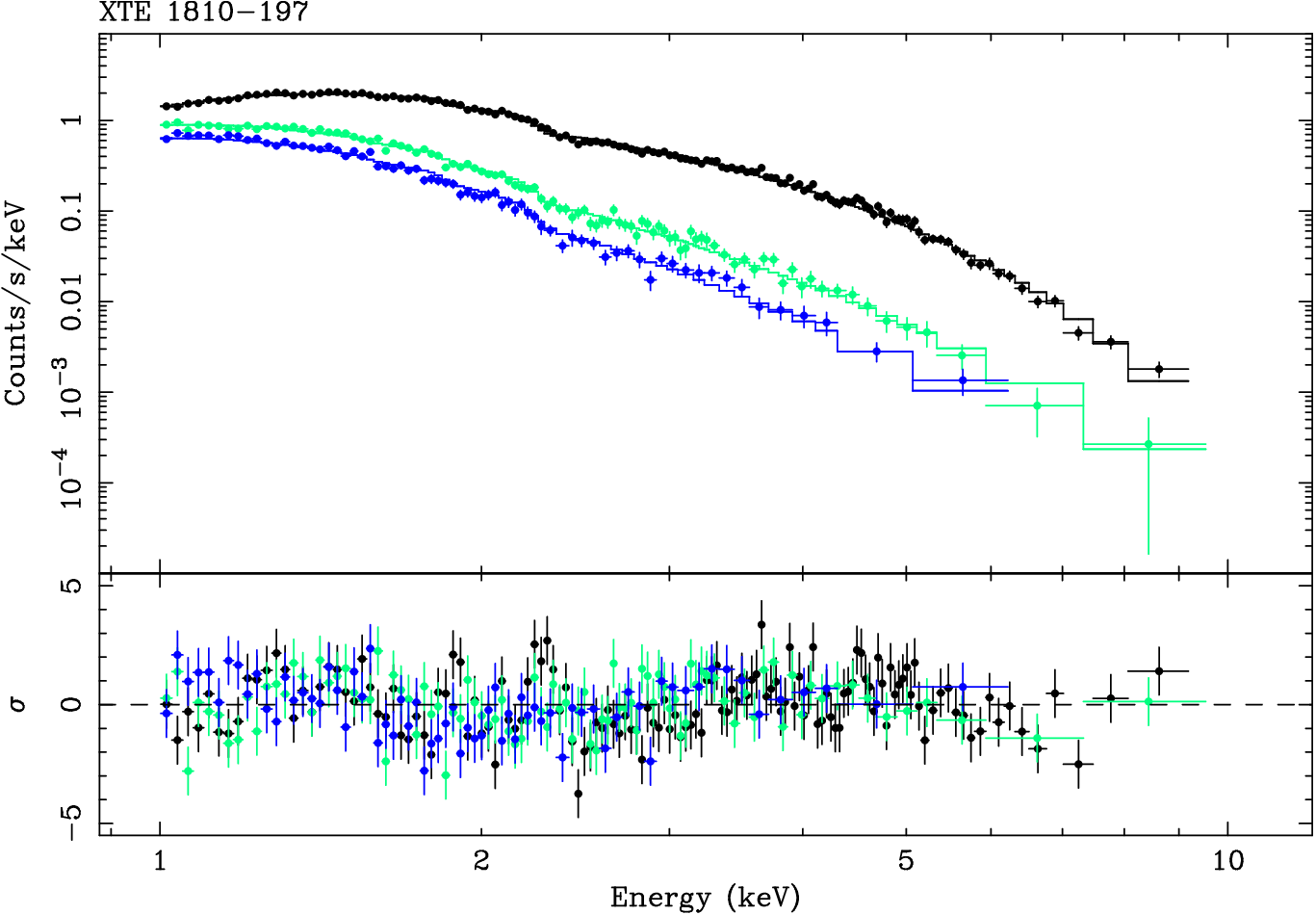,height=5cm,angle=270,width=7cm}
\psfig{figure=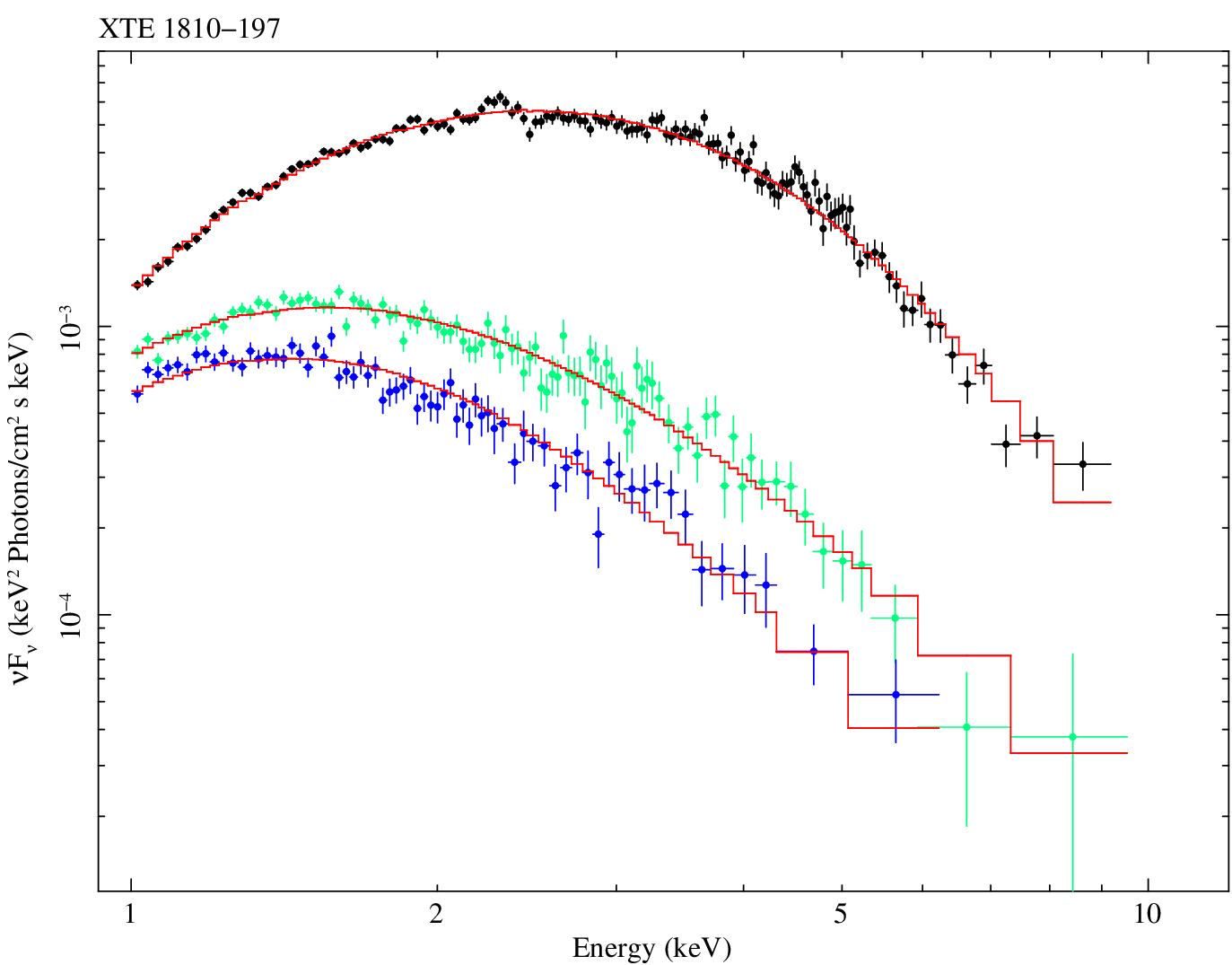,height=5cm,angle=270,width=7cm}}}}
\caption{\xte: left column represents the spectra in Counts/s/keV while in the
right column we report the $\nu$F$_\nu$ plots. The upper panels are
relative to the modeling with two absorbed blackbodies (BB+BB), while
bottom panels report the resonant cyclotron scattering model
(RCS). See also Tab.\,\ref{table1810} and \S\,\ref{res:trans} for
details. Black, light-green and blue colors are relative to
observations taken on 2004, 2005 and 2006, respectively. The red lines
represent the total model, while the dashed lines are the single
components.}
\label{spectra1810}
\end{figure*}


\begin{table*}
\setlength{\tabcolsep}{0.01in}
\centering
\caption{Spectral Parameters: \xte}
\begin{tabular}{lcccccc}
\hline
\hline
 {\em AXP} & \multicolumn{6}{c}{\xte} \\
\hline
 & \multicolumn{2}{c}{2004}  & \multicolumn{2}{c}{2005} &  \multicolumn{2}{c}{2006} \\
\hline
\multicolumn{1}{l}{Parameters} & BB+BB & RCS & BB+BB & RCS & BB+BB  &  RCS \\
\hline

N$_{H}$
& 0.58$^{+0.06}_{-0.05}$ & 0.40$^{+0.05}_{-0.05}$
& 0.52$^{+0.08}_{-0.07}$ & 0.25$^{+0.05}_{-0.05}$
& 0.4$^{+0.1}_{-0.1}$ & 0.14$^{+0.22}_{-0.05}$ \\
 & & & & & &   \\

kT$_1$ (keV)
& 0.36$^{+0.02}_{-0.02}$ &  0.44$^{+0.03}_{-0.03}$
& 0.27$^{+0.03}_{-0.02}$ & 0.29$^{+0.08}_{-0.07}$
& 0.25$^{+0.03}_{-0.04}$ & 0.13$^{+0.05}_{-0.05}$ \\

BB$_1$~norm
& 6.6$^{+0.5}_{-0.4}\times10^{-5}$ &
& 3.8$^{+0.2}_{-0.1}\times10^{-5}$  &
& 2.7$^{+0.3}_{-0.3}\times10^{-5}$ & \\

kT$_2$ (keV)
& 0.71$^{+0.01}_{-0.02}$ &
& 0.58$^{+0.03}_{-0.03}$  &
& 0.36$^{+0.05}_{-0.07}$ & \\

BB$_2$~norm
& 12$^{+1}_{-1}\times10^{-5}$  &
& 1.5$^{+0.1}_{-0.1}\times10^{-5}$ &
& 0.7$^{+0.1}_{-0.1}\times10^{-5}$ & \\
 & & & & & & \\

$\beta_{T}$
& & 0.19$^{+0.05}_{-0.05}$
&  & 0.40$^{+0.05}_{-0.05}$
&  & 0.35$^{+0.05}_{-0.05}$ \\

$\tau_{res}$
& & 5.9$^{+1.6}_{-1.0}$
&  &  1.6$^{+0.2}_{-0.2}$
& & 1.4$^{+0.1}_{-0.1}$ \\

RCS~norm
& & 1.1$^{+0.3}_{-0.3}\times10^{-4}$
&    & 7.2$^{+0.3}_{-0.4}\times10^{-5}$
& & 2.5$^{+0.5}_{-0.5}\times10^{-4}$\\
 & & & & & &  \\

Flux 1--10\,keV
& 12$^{+3}_{-2}\times10^{-12}$ &   11$^{+3}_{-3}\times10^{-12}$
& 2.2$^{+0.1}_{-0.2}\times10^{-12}$ & 2.1$^{+0.2}_{-0.2}\times10^{-12}$
& 1.2$^{+0.3}_{-0.4}\times10^{-12}$ & 1.2$^{+0.5}_{-0.4}\times10^{-12}$\\

 & & & & & &  \\
$\chi^2_{\nu}$ (dof)
& 1.21 (135) &  1.27 (135)
&   0.94 (97) & 1.07 (97)
& 0.97 (67) & 1.00 (67) \\

\hline
\hline
\end{tabular}
\tablecomments{Best fit values of the spectral parameters obtained
by fitting several $\sim$1--10\,keV \XMM\, spectra, taken in different
source states, with two absorbed blackbodies (BB+BB), and with a
resonant cyclotron scattering model (RCS). Errors are at 1$\sigma$
confidence level, reported fluxes are absorbed and in units of
\ergscm2 , and N$_{H}$ in units of $10^{22}$\,cm$^{-2}$ and assuming
solar abundances from Lodders (2003); 2\% systematic error has been
included. See also Fig.\,\ref{spectra1810} and \S\,\ref{res:trans} for
details.}
\label{table1810}
\end{table*}



\begin{figure*}
\centering{
\hspace{0.1cm}
\vbox{
\hbox{
\psfig{figure=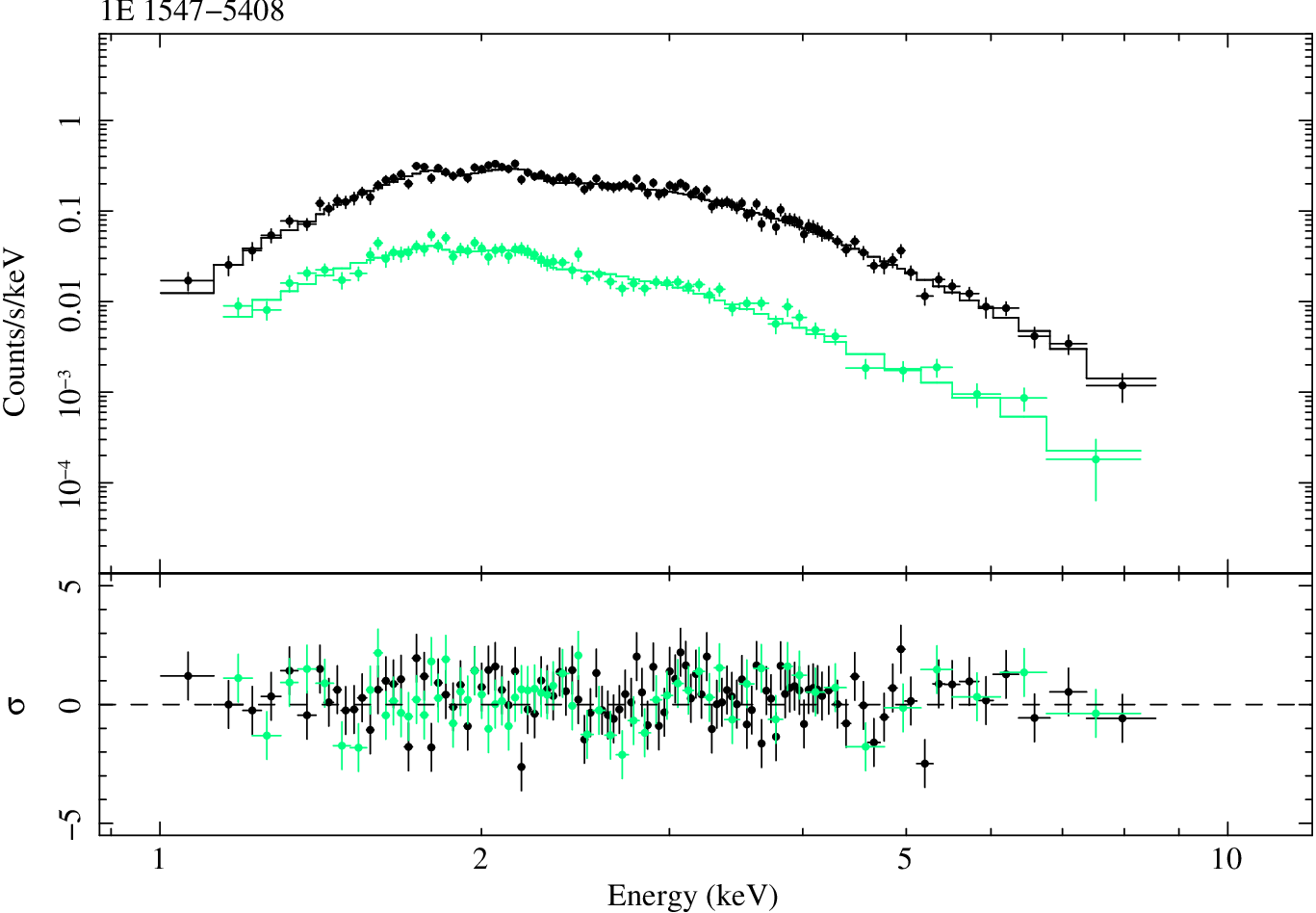,height=5cm,angle=270,width=7cm}
\psfig{figure=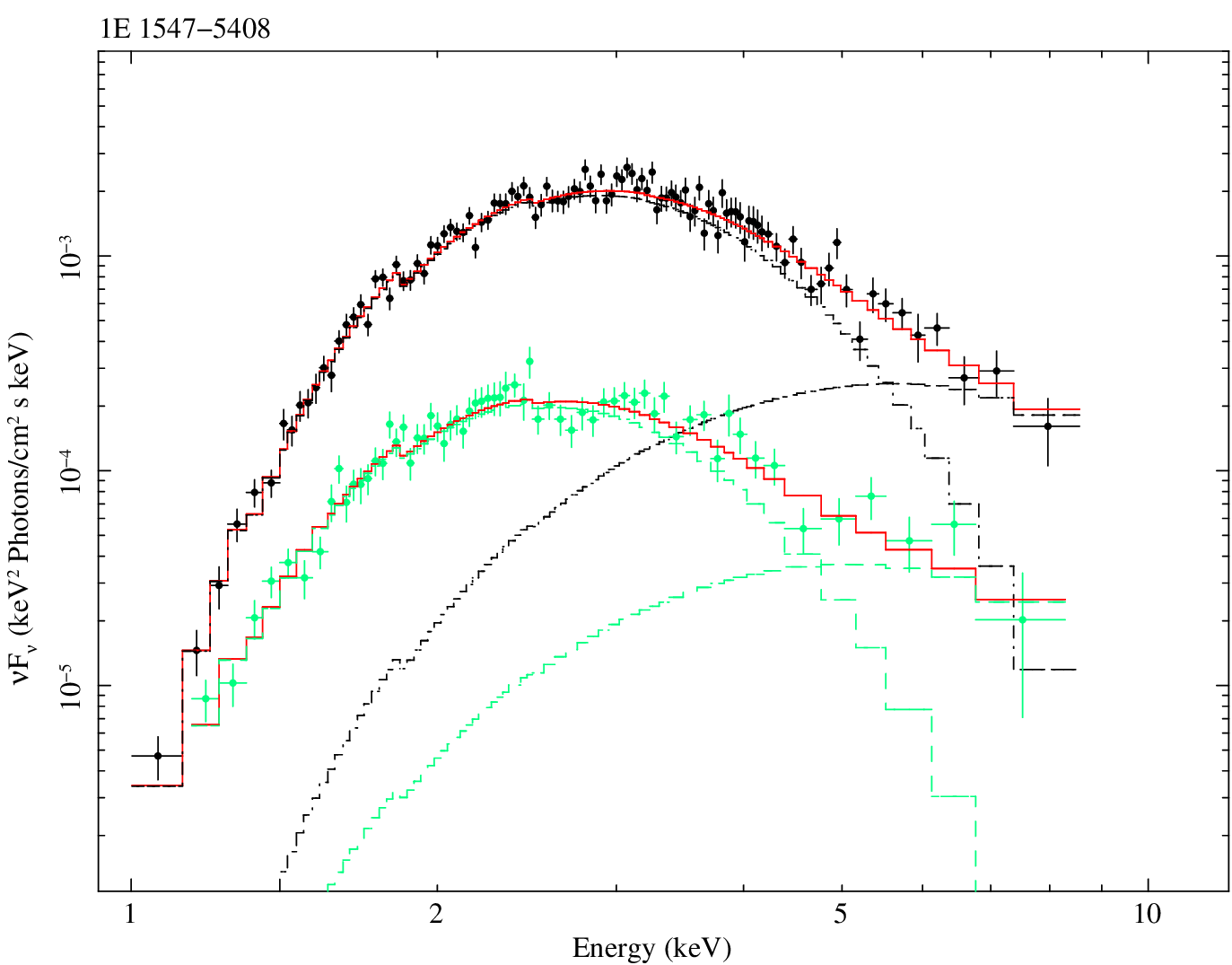,height=5cm,angle=270,width=7cm}}
\hbox{
\psfig{figure=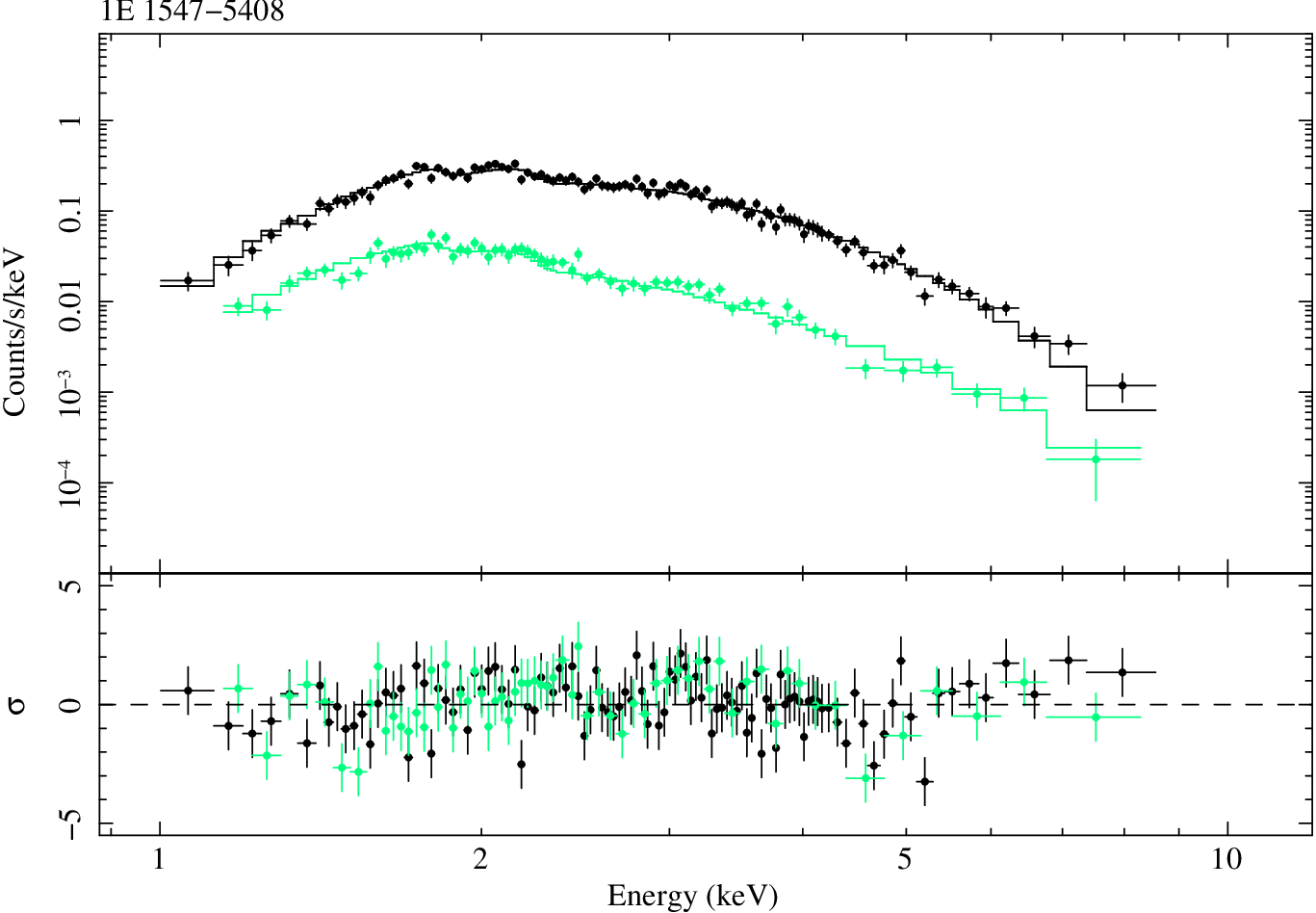,height=5cm,angle=270,width=7cm}
\psfig{figure=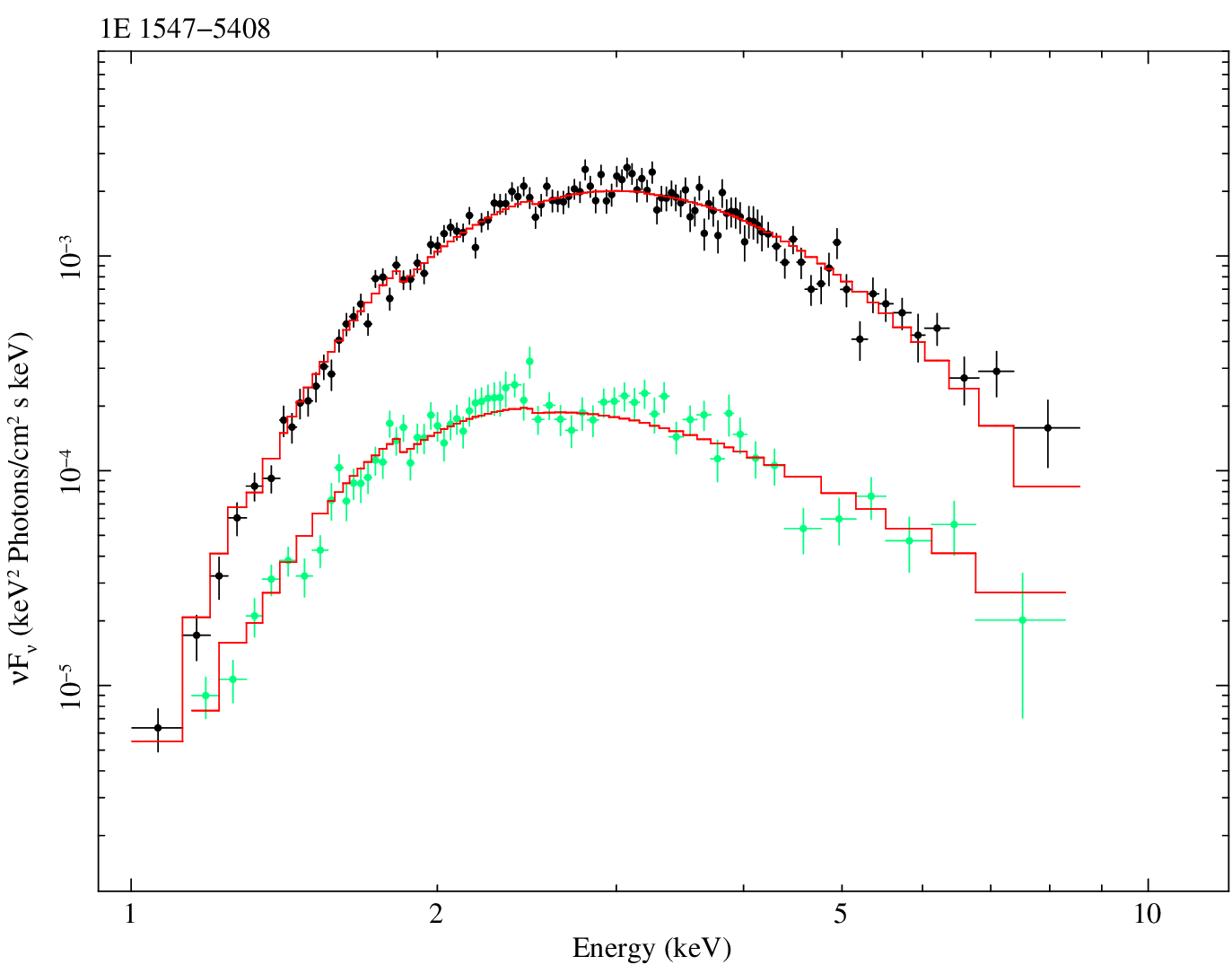,height=5cm,angle=270,width=7cm}}}}
\caption{\1e: left column represents the spectra in Counts/s/keV while in the
right column we report the $\nu$F$_\nu$ plots. The upper panels are
relative to the modeling with two blackbodies (BB+BB), while bottom
panels report the resonant cyclotron scattering model (RCS). See also
Tab.\,\ref{table1547} and \S\,\ref{res:trans} for details. Black and
light-green colors are relative to observations taken on 2007 and
2006, respectively. The red lines represent the total model, while the
dashed lines are the single components.}
\label{spectra1547}
\end{figure*}

\begin{table*}
\setlength{\tabcolsep}{0.01in}
\centering
\caption{Spectral Parameters: \1e}
\begin{tabular}{lcccc}
\hline
\hline
 {\em AXP} & \multicolumn{4}{c}{\1e} \\
\hline
 & \multicolumn{2}{c}{2006}
&  \multicolumn{2}{c}{2007} \\
\hline

\multicolumn{1}{l}{Parameters}
& BB+BB & RCS
& BB+BB & RCS  \\
\hline

N$_{H}$
& 3.76$^{+0.06}_{-0.05}$ & 2.8$^{+0.1}_{-0.1}$
& 4.58$^{+0.08}_{-0.07}$ & 4.6$^{+0.1}_{-0.1}$  \\
 & & & &   \\

kT$_1$ (keV)
& 0.46$^{+0.03}_{-0.02}$ &  0.33$^{+0.05}_{-0.05}$
& 0.51$^{+0.02}_{-0.02}$ & 0.46$^{+0.08}_{-0.05}$ \\

BB$_1$~norm
&  1.2$^{+0.5}_{-0.4}\times10^{-5}$ &
& 7.2$^{+0.5}_{-0.5}\times10^{-6}$  & \\

kT$_2$ (keV)
& 1.2$^{+0.1}_{-0.1}$ &
& 1.34$^{+0.08}_{-0.07}$  &  \\

BB$_2$~norm
& 1.4$^{+0.1}_{-0.1}\times10^{-6}$  &
& 1.4$^{+0.1}_{-0.1}\times10^{-4}$ &  \\
 & & & & \\

$\beta_{T}$
& & 0.32$^{+0.03}_{-0.09}$
&  & 0.24$^{+0.04}_{-0.04}$ \\

$\tau_{res}$
& & 1.0$^{+0.8}_{-0.2}$
&  &  1.0$^{+0.1}_{-0.1}$ \\

RCS~norm
& & 2.6$^{+0.3}_{-0.3}\times10^{-5}$
&  & 9.4$^{+0.3}_{-0.4}\times10^{-5}$  \\
 & & & &  \\

Flux 1--10\,keV
& 3.2$^{+0.1}_{-0.1}\times10^{-13}$  &   3.1$^{+0.1}_{-0.2}\times10^{-13}$
& 3.0$^{+0.1}_{-0.1}\times10^{-12}$ & 3.0$^{+0.2}_{-0.2}\times10^{-12}$ \\
 & & & &  \\

$\chi^2_{\nu}$ (dof)
& 1.18 (60) &  1.20 (60)
&   1.02 (105) & 1.13 (105)  \\

\hline
\hline
\end{tabular}
\tablecomments{Best fit values of the spectral parameters obtained
by fitting several $\sim$1--10\,keV \XMM\, spectra, taken in different
source states, with two absorbed blackbodies (BB+BB), and with a
resonant cyclotron scattering model (RCS). Errors are at 1$\sigma$
confidence level, reported fluxes are absorbed and in units of
\ergscm2 , and N$_{H}$ in units of $10^{22}$\,cm$^{-2}$ and assuming
solar abundances from Lodders (2003); 2\% systematic error has been
included. See also Fig.\,\ref{spectra1547} and \S\,\ref{res:trans} for
details.}
\label{table1547}
\end{table*}


\begin{figure*}
\centering{
\hspace{0.1cm}
\vbox{
\hbox{
\psfig{figure=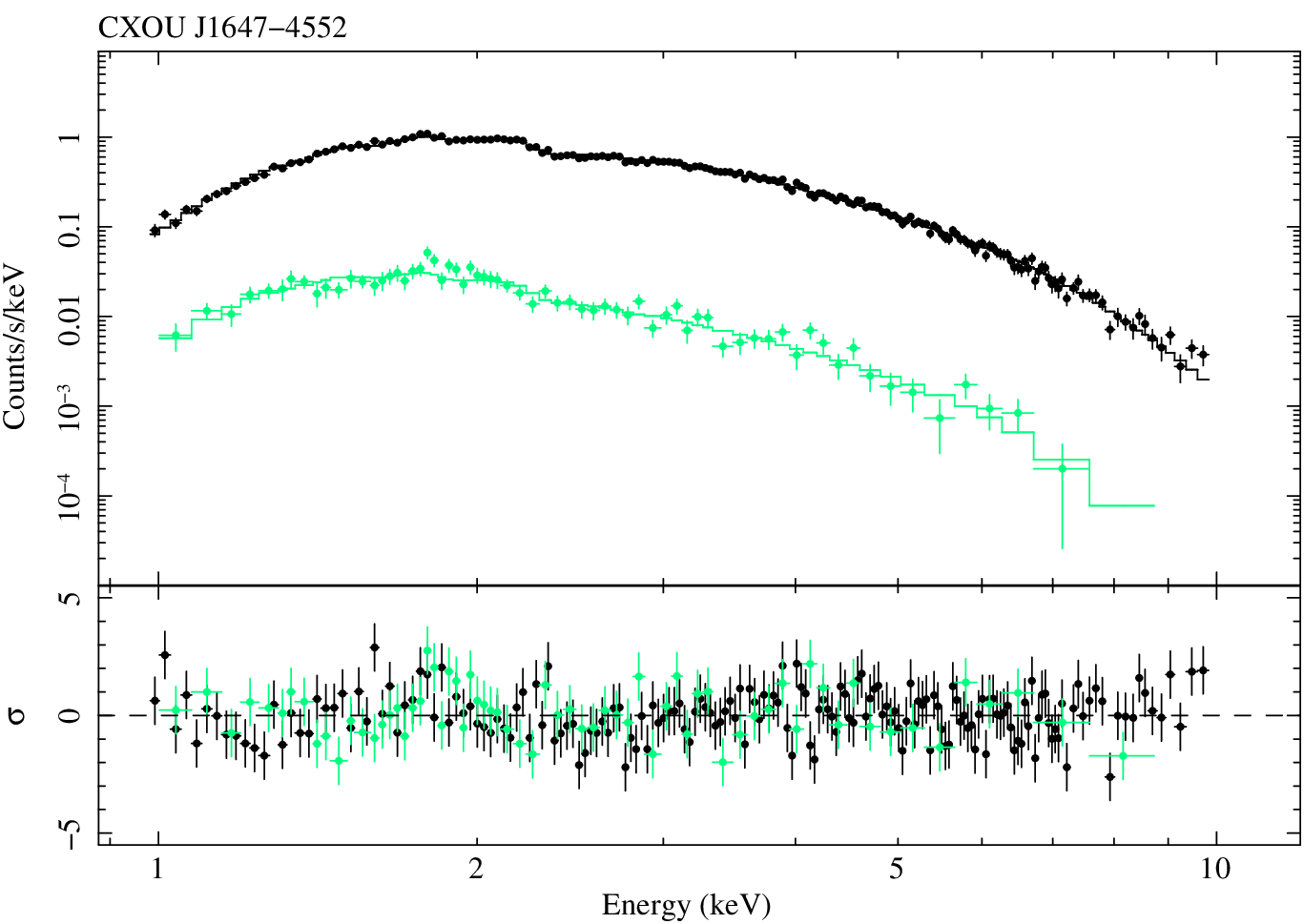,height=5cm,angle=270,width=7cm}
\psfig{figure=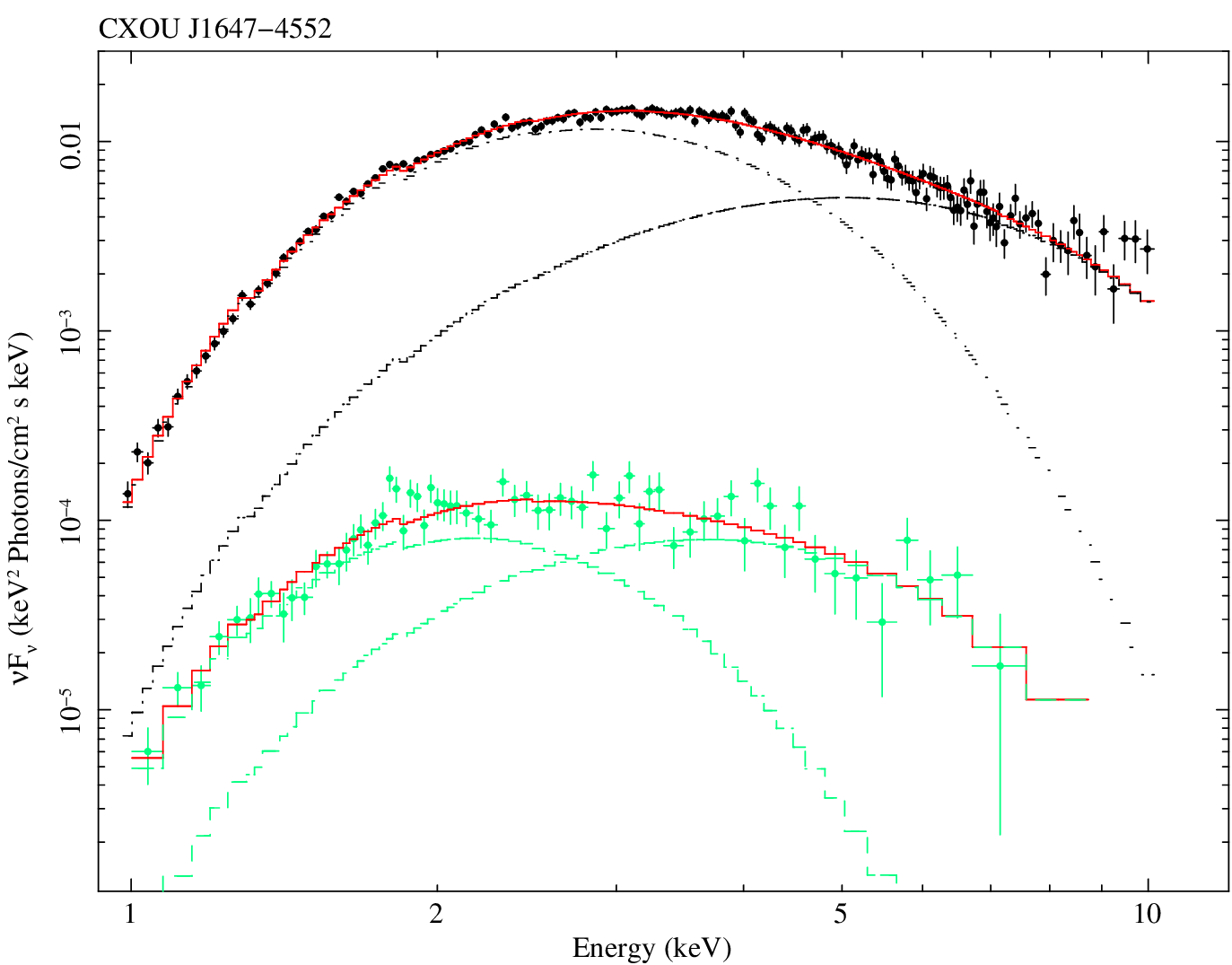,height=5cm,angle=270,width=7cm}}
\hbox{
\psfig{figure=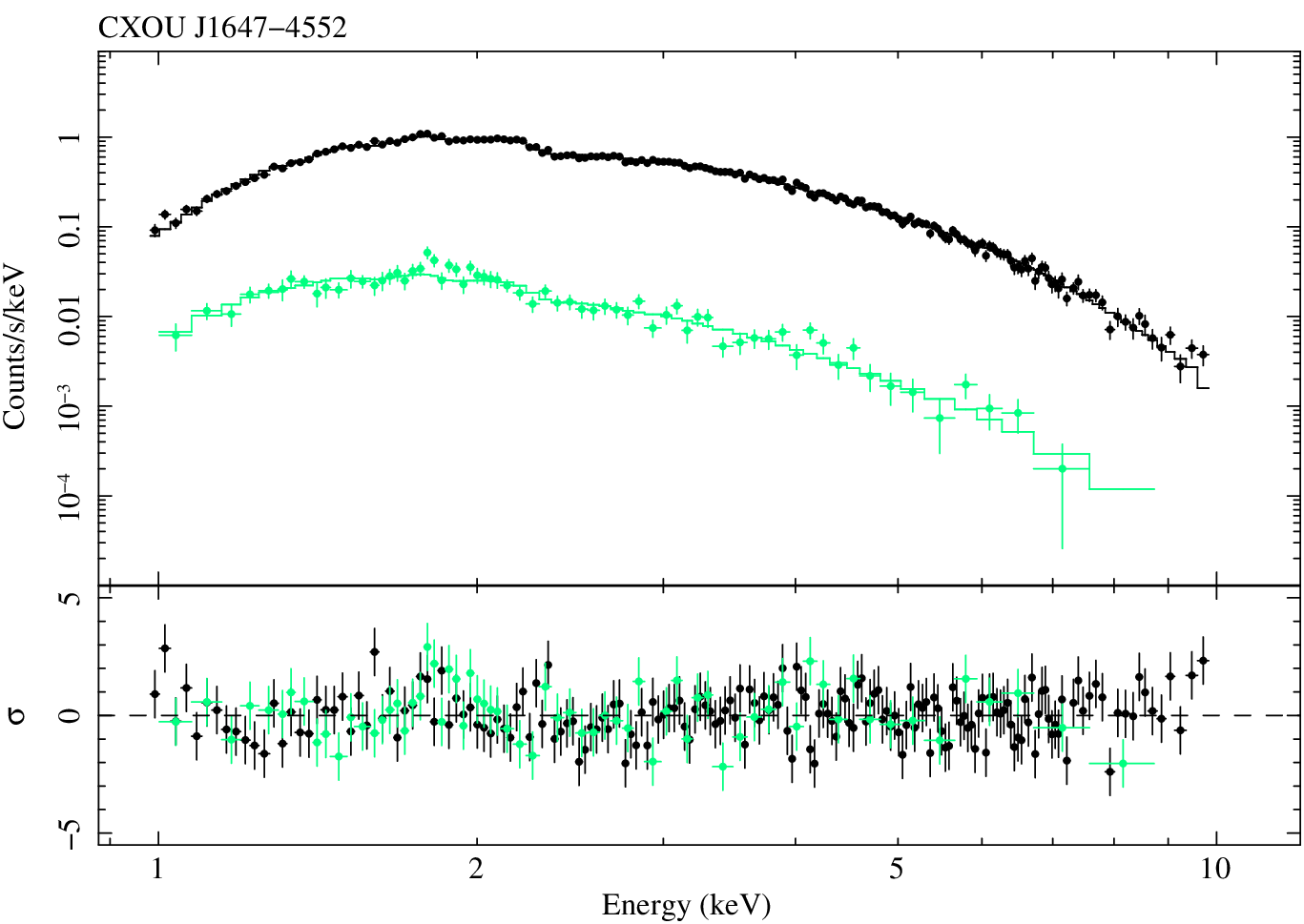,height=5cm,angle=270,width=7cm}
\psfig{figure=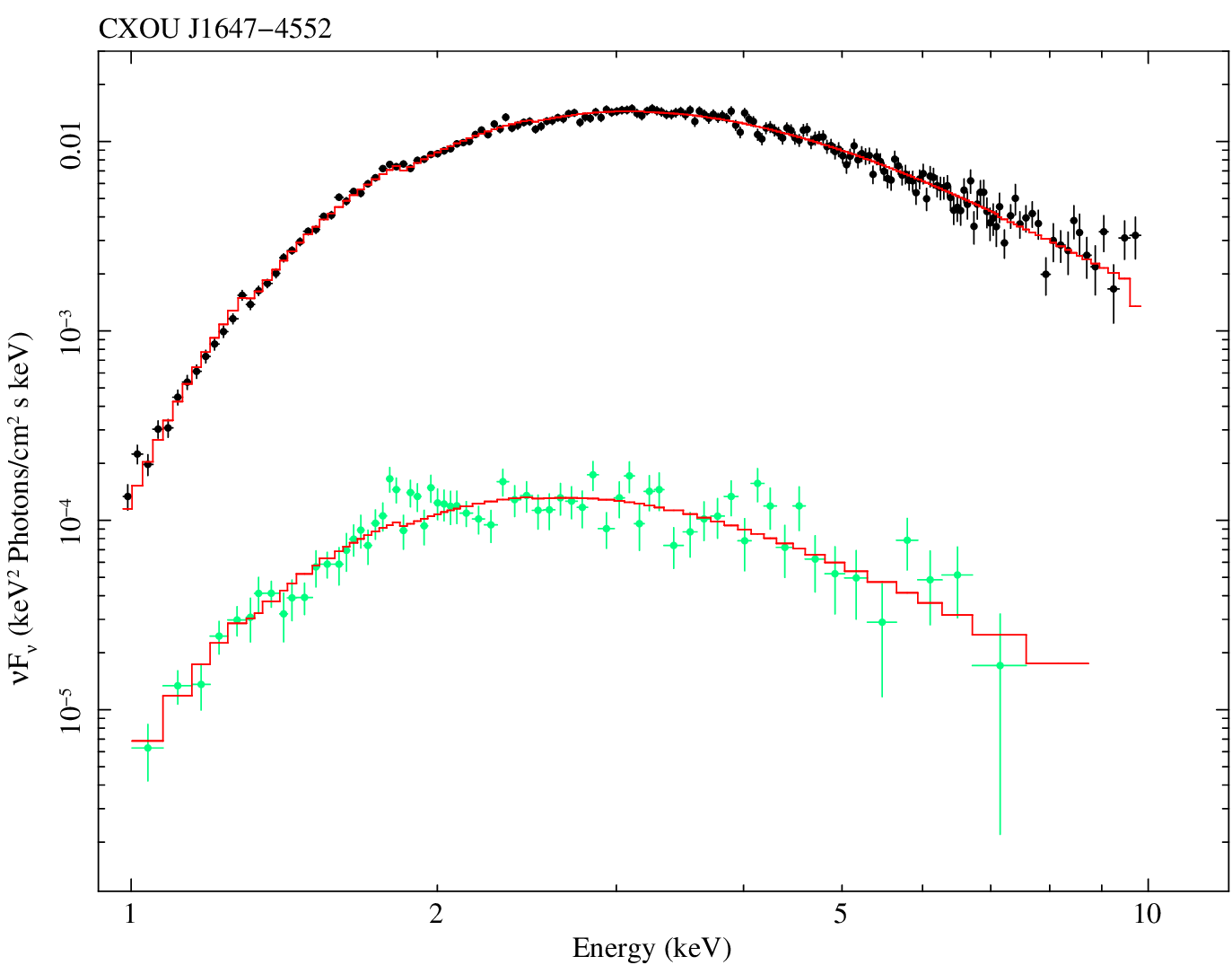,height=5cm,angle=270,width=7cm}}}}
\caption{\wes: left column represents the spectra in Counts/s/keV while in the right
column we report the $\nu$F$_\nu$ plots. The upper panels are relative
to the modeling with two absorbed blackbodies (BB+BB), while bottom
panels report the resonant cyclotron scattering model (RCS). See also
Tab.\,\ref{table1647} and \S\,\ref{res:trans} for details. Black and
light-green colors are relative to observations taken on 2006
September 22 and 16, respectively. The red lines represent the total
model, while the dashed lines are the single components.}
\label{spectra1647}
\end{figure*}

\begin{table*}
\setlength{\tabcolsep}{0.01in}
\centering
\caption{Spectral Parameters: \wes}
\begin{tabular}{lcccc}
\hline
\hline
 {\em AXP}  & \multicolumn{4}{c}{\wes} \\
\hline
 & \multicolumn{2}{c}{2006/09/16} &
\multicolumn{2}{c}{2006/09/22} \\
\hline

\multicolumn{1}{l}{Parameters}
& BB+BB & RCS
& BB+BB & RCS  \\
\hline

N$_{H}$
& 2.14$^{+0.06}_{-0.06}$ & 2.08$^{+0.15}_{-0.16}$
& 2.34$^{+0.04}_{-0.04}$ & 2.40$^{+0.04}_{-0.04}$  \\
 & & & &  \\

kT$_1$ (keV)
& 0.39$^{+0.03}_{-0.02}$ &  0.34$^{+0.15}_{-0.19}$
& 0.59$^{+0.02}_{-0.02}$ & 0.55$^{+0.08}_{-0.08}$ \\

BB$_1$~norm
&  4.5$^{+0.5}_{-0.4}\times10^{-6}$ &
& 4.4$^{+0.5}_{-0.5}\times10^{-4}$  & \\

kT$_2$ (keV)
& 0.85$^{+0.1}_{-0.1}$ &
& 1.23$^{+0.04}_{-0.04}$  &  \\

BB$_2$~norm
& 2.4$^{+0.1}_{-0.1}\times10^{-6}$  &
& 1.4$^{+0.1}_{-0.1}\times10^{-4}$ &  \\
 & & & & \\

$\beta_{T}$
& & 0.30$^{+0.08}_{-0.08}$
&   & 0.42$^{+0.08}_{-0.08}$ \\

$\tau_{res}$
& & 2.9$^{+0.1}_{-0.1}$
&  &  1.09$^{+0.05}_{-0.05}$ \\

RCS~norm
& & 7.8$^{+0.3}_{-0.3}\times10^{-6}$
&  & 3.0$^{+0.3}_{-0.4}\times10^{-3}$ \\
 & & & &  \\

Flux 1--10\,keV
& 2.4$^{+0.1}_{-0.1}\times10^{-13}$ &  2.4$^{+0.1}_{-0.1}\times10^{-13}$
& 2.2$^{+0.1}_{-0.1}\times10^{-11}$ & 2.2$^{+0.1}_{-0.1}\times10^{-11}$ \\
 & & & &  \\

$\chi^2_{\nu}$ (dof)
& 1.00 (73) & 1.23  (73)
&   1.01 (136) & 1.06 (136)  \\

\hline
\hline
\end{tabular}
\tablecomments{Best fit values of the spectral parameters obtained
by fitting several $\sim$1--10\,keV \XMM\, spectra, taken in different
source states, with two absorbed blackbodies (BB+BB), and with a
resonant cyclotron scattering model (RCS). Errors are at 1$\sigma$
confidence level, reported fluxes are absorbed and in units of
\ergscm2 , and N$_{H}$ in units of $10^{22}$\,cm$^{-2}$ and assuming
solar abundances from Lodders (2003); 2\% systematic error has been
included. See also Fig.\,\ref{spectra1647} and \S\,\ref{res:trans} for
details.}
\label{table1647}
\end{table*}



\begin{figure*}
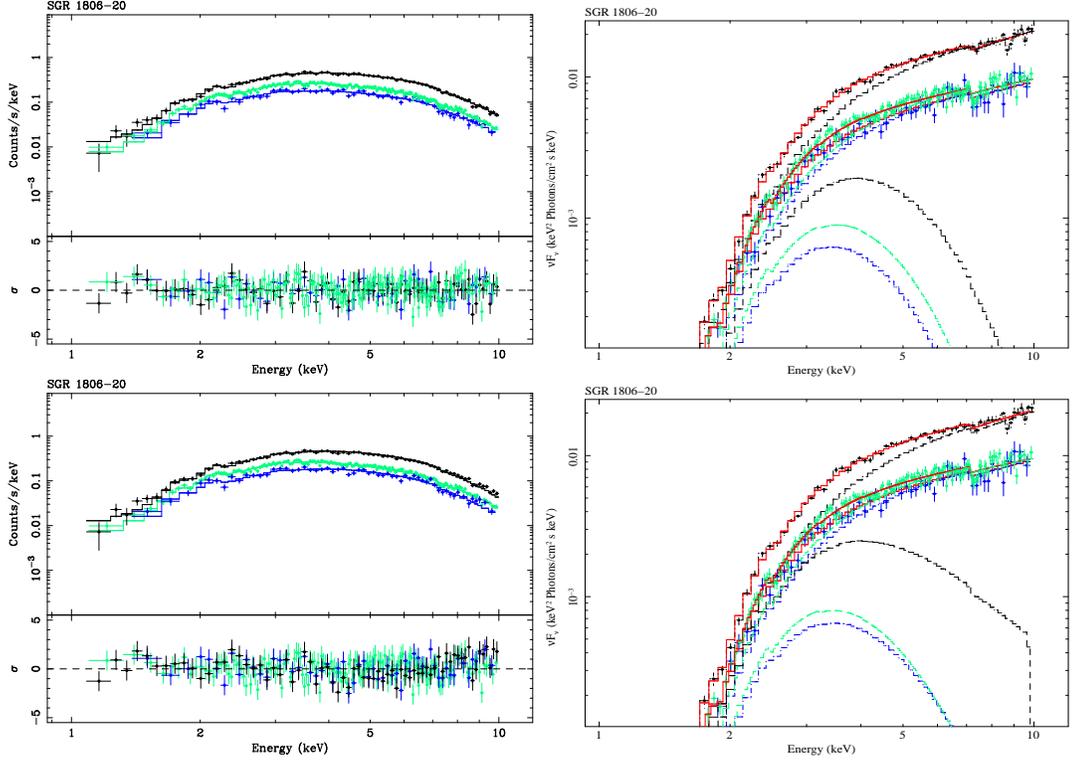

\centering{
\hspace{0.1cm}
\vbox{
\hbox{
\psfig{figure=f8a.eps,height=5cm,angle=270,width=7cm}
\psfig{figure=f8b.eps,height=5cm,angle=270,width=7cm}}
\hbox{
\psfig{figure=f8c.eps,height=5cm,angle=270,width=7cm}
\psfig{figure=f8d.eps,height=5cm,angle=270,width=7cm}}}}
\caption{\sgra: left column shows the spectra in Counts/s/keV while in the right
column we report the $\nu$F$_\nu$ plots. The upper panels are relative
to the modeling with a blackbody plus power-law (BB+PL), while bottom
panels report the resonant cyclotron scattering model plus power-law
(RCS+PL). See also Tab.\,\ref{table1806} and \S\ref{res:sgrs} for
details. Light green, black and blue colours are relative to
observations taken on 2003, 2004 and 2005, respectively. The red lines
represent the total model, while the dashed lines are the single
components.}
\label{spectra1806}
\end{figure*}


\begin{table*}
\setlength{\tabcolsep}{0.01in}
\centering
\caption{Spectral Parameters: \sgra}
\begin{tabular}{lcccccc}
\hline
\hline
 {\em SGR} & \multicolumn{6}{c}{\sgra} \\
\hline
 &  \multicolumn{2}{c}{2003}
 & \multicolumn{2}{c}{2004}
& \multicolumn{2}{c}{2005} \\
\hline
 \multicolumn{1}{l}{Parameters}
& BB+PL & RCS+PL
& BB+PL & RCS+PL
& BB+PL & RCS+PL  \\
\hline

  N$_{H}$
& 9.9$^{+0.4}_{-0.4}$  & 9.3$^{+1.0}_{-0.8}$
&   9.7$^{+0.2}_{-0.2}$ &  10.1$^{+0.6}_{-0.8}$
& 10.2$^{+1.0}_{-0.8}$  & 11.0$^{+1.0}_{-1.0}$ \\
& & & & & &   \\

kT (keV)
&  0.56$^{+0.05}_{-0.04}$ &  0.57$^{+0.06}_{-0.1}$
&  0.72$^{+0.06}_{-0.07}$ & 0.54$^{+0.06}_{-0.05}$
& 0.57$^{+0.04}_{-0.04}$ & 0.26$^{+0.07}_{-0.08}$ \\

BB~norm
& 5.5$^{+0.3}_{-0.3}\times10^{-5}$ &
& 1.0$^{+0.4}_{-0.3}\times10^{-4}$  &
& 7.4$^{+0.4}_{-0.3}\times10^{-5}$ & \\
& & & & & &  \\

$\beta_{T}$
& & 0.17$^{+0.03}_{-0.03}$
& & 0.14$^{+0.08}_{-0.03}$
& & 0.49$^{+0.04}_{-0.03}$   \\

$\tau_{res}$
&  & 2.2$^{+1.5}_{-1.1}$
&  & 4.3$^{+0.7}_{-1.1}$
& & 2.6$^{+0.2}_{-0.3}$ \\

RCS~norm
& & 3.8$^{+0.5}_{-0.5}\times10^{-5}$
& & 7.4$^{+0.7}_{-0.8}\times10^{-5}$
&  & 4.6$^{+0.7}_{-0.8}\times10^{-4}$\\
& & & & & &  \\

$\Gamma$
& 1.5$^{+0.1}_{-0.1}$ & 1.2$^{+0.1}_{-0.1}$
&  1.3$^{+0.1}_{-0.1}$ & 1.3$^{+0.1}_{-0.1}$
& 1.5$^{+0.1}_{-0.1}$ & 1.2$^{+0.2}_{-0.1}$ \\

PL$$~norm
& 3.1$^{+0.2}_{-0.2}\times10^{-3}$ & 1.7$^{+0.2}_{-0.3}\times10^{-3}$
& 4.7$^{+0.2}_{-0.2}\times10^{-3}$  & 5.1$^{+0.4}_{-0.3}\times10^{-3}$
& 3.8$^{+0.2}_{-0.3}\times10^{-3}$  &   1.7$^{+0.8}_{-1.0}\times10^{-3}$  \\
 & & & &  &  & \\

Flux 1--10\,keV
&   1.2$^{+0.5}_{-0.6}\times10^{-11}$ & 1.2$^{+0.8}_{-0.8}\times10^{-11}$
& 2.6$^{+0.6}_{-0.7}\times10^{-11}$ & 2.6$^{+0.7}_{-0.8}\times10^{-11}$
& 1.4$^{+0.5}_{-0.6}\times10^{-11}$  & 1.3$^{+0.8}_{-0.8}\times10^{-11}$ \\

 & & & & & & \\
$\chi^2_{\nu}$ (dof)
&  0.96 (54) &  1.03 (52)
& 1.01 (65) &   0.97 (63)
&  1.02 (159) & 0.90 (157) \\

\hline
\hline
\end{tabular}
\tablecomments{Best fit values of the spectral parameters obtained
by fitting several $\sim$1--10\,keV \XMM\, spectra, taken in different
source states, with a blackbody plus power-law model (BB+PL), and with
a resonant cyclotron scattering plus power-law model (RCS+PL). Errors
are at 1$\sigma$ confidence level, reported fluxes are absorbed and in
units of \ergscm2 , and N$_{H}$ in units of $10^{22}$\,cm$^{-2}$ and
assuming solar abundances from Lodders (2003); 2\% systematic error
has been included. See also Fig.\,\ref{spectra1806} and
\S\ref{res:sgrs} for details.}
\label{table1806}
\end{table*}



\begin{figure*}
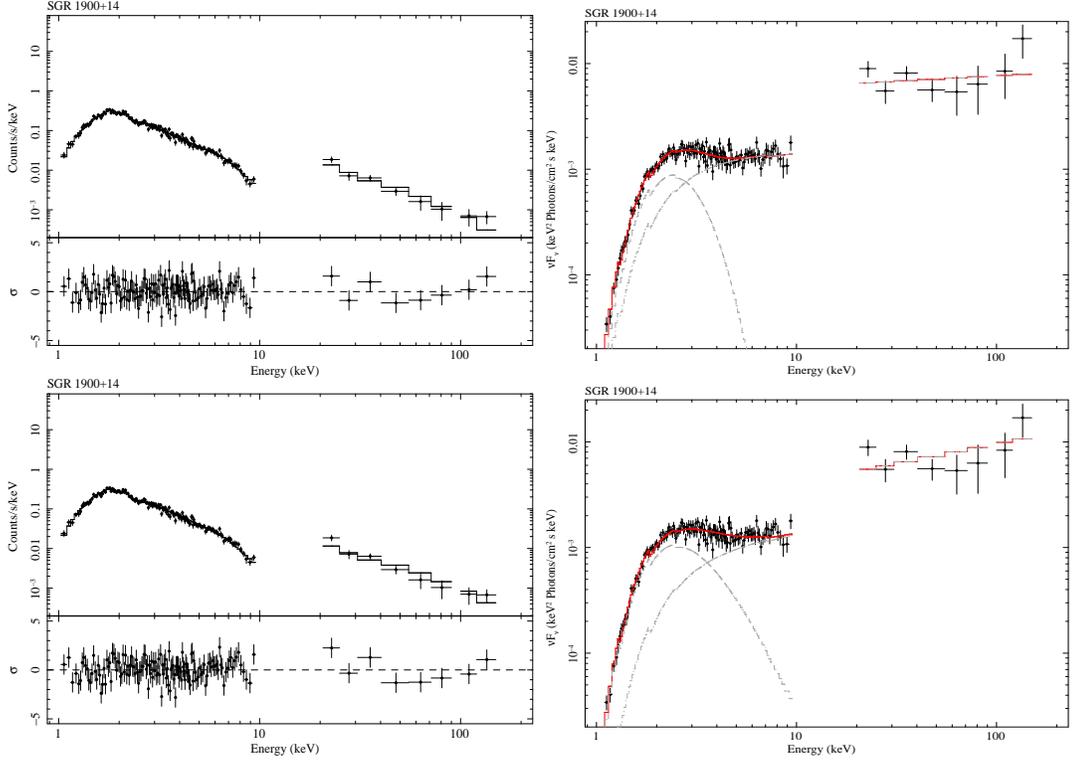

\centering{
\hspace{0.1cm}
\vbox{
\hbox{
\psfig{figure=f9a.eps,height=5cm,width=7cm,angle=270}
\psfig{figure=f9b.eps,height=5cm,width=7cm,angle=270}}
\hbox{
\psfig{figure=f9c.eps,height=5cm,width=7cm,angle=270}
\psfig{figure=f9d.eps,height=5cm,width=7cm,angle=270}}}}
\caption{\sgrb: left column shows the spectra in Counts/s/keV while in the right
column we report the $\nu$F$_\nu$ plots. The upper panels are relative
to the modeling with a blackbody plus power-law (BB+PL), while bottom
panels report the resonant cyclotron scattering model plus power-law
(RCS+PL). See also Tab.\,\ref{table1900} and \S\ref{res:sgrs} for
details. The red lines represent the total model, while the dashed
lines are the single components.}
\label{spectra1900}
\end{figure*}


\begin{table*}
\setlength{\tabcolsep}{0.01in}
\centering
\caption{Spectral Parameters: \sgrb}
\begin{tabular}{lcc}
\hline
\hline
 {\em SGR} & \multicolumn{2}{c}{SGR\,1900+14} \\
\hline
\multicolumn{1}{l}{Parameters}
& BB+PL & RCS+PL \\
\hline

N$_{H}$
& 3.5$^{+0.1}_{-0.1}$ & 4.0$^{+0.1}_{-0.1}$ \\

constant & 1.20 & 1.10 \\
 & &    \\

kT (keV)
& 0.45$^{+0.04}_{-0.04}$ & 0.30$^{+0.08}_{-0.1}$ \\

BB~norm
& 6.7$^{+0.1}_{-0.1}\times10^{-5}$ &  \\
 & &    \\

$\beta_{T}$
& & 0.26$^{+0.03}_{-0.03}$ \\

$\tau_{res}$
& &  2.5$^{+0.5}_{-0.2}$ \\

RCS~norm
& &  1.8$^{+0.04}_{-0.05}\times10^{-4}$ \\
 & &    \\

$\Gamma$
& 1.4$^{+0.1}_{-0.1}$  & 1.24$^{+0.07}_{-0.07}$  \\

PL~norm
& 4.4$^{+0.1}_{-0.1}\times10^{-4}$ & 3.0$^{+0.1}_{-0.1}\times10^{-4}$\\
 & &  \\

Flux 1--10\,keV
& 3.9$^{+0.1}_{-0.1}\times10^{-12}$ & 3.8$^{+0.1}_{-0.1}\times10^{-12}$  \\

Flux 1--200\,keV
& 1.7$^{+0.1}_{-0.1}\times10^{-11}$ & 1.7$^{+0.1}_{-0.1}\times10^{-11}$  \\
& &  \\

$\chi^2_{\nu}$ (dof)
& 1.18 (141)  &   1.15 (139)  \\

\hline
\hline
\end{tabular}
\tablecomments{Best fit values of the spectral parameters obtained
by fitting the $\sim$1--200\,keV \XMM\, and \INT\, spectra with a
blackbody plus a power-law model (BB+PL), and with a resonant
cyclotron scattering model plus a power-law (RCS+PL). Errors are at
1$\sigma$ confidence level, reported fluxes are absorbed and in units
of \ergscm2 , and N$_{H}$ in units of $10^{22}$\,cm$^{-2}$ and
assuming solar abundances from Lodders (2003); 2\% systematic error
has been included. See also Fig.\,\ref{spectra1900} and
\S\ref{res:sgrs} for details.}
\label{table1900}
\end{table*}


\newpage


\begin{figure*}
\hbox{
\hspace{4.5cm}
\psfig{figure=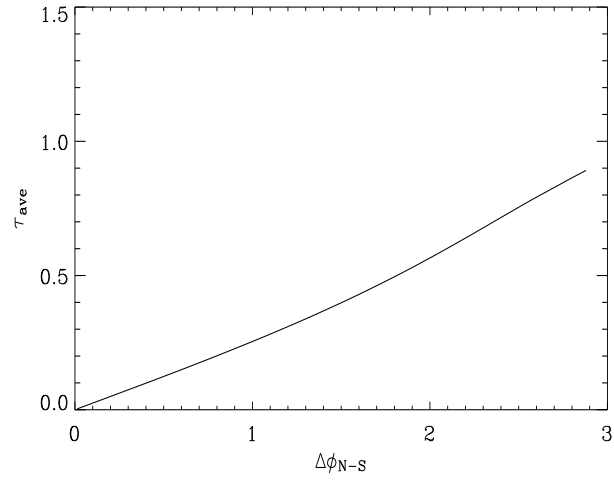,height=7cm,width=9cm,angle=0}}
\caption{Angle-averaged optical depth  in a twisted
magnetosphere model (Thompson Lyutikov \& Kulkarni 2002) as a function
of the twist angle. The curve refers to $\beta_{bulk}=1$; for
different values of the bulk velocity the ordinate should be divided
by $\beta_{bulk}$. }
\label{taures}
\end{figure*}



\begin{figure*}
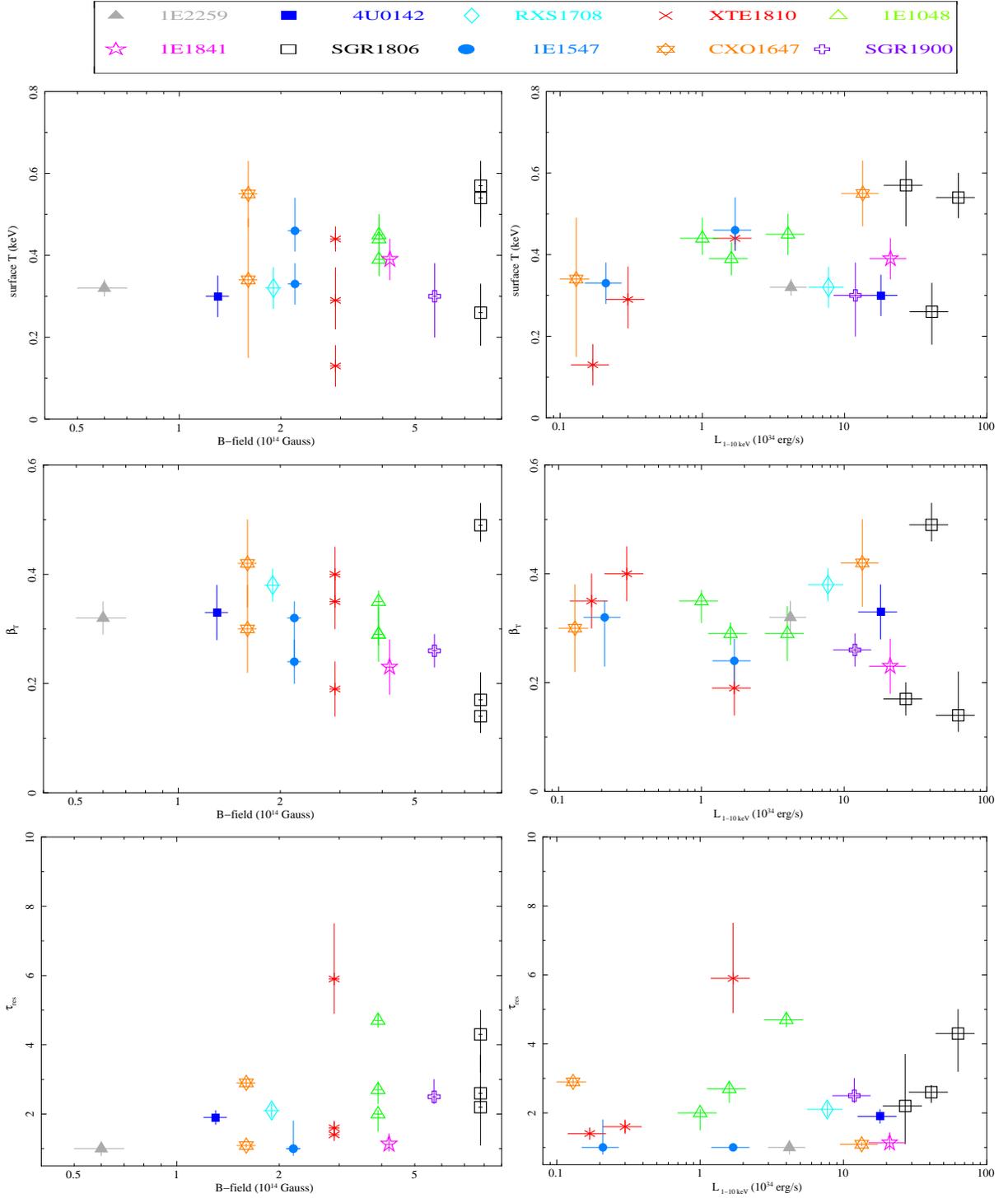

\centering{
\hspace{0.1cm}
\vbox{
\hspace{0.5cm}
\hbox{
\psfig{figure=f11a.eps,height=1.2cm,angle=270,width=14cm} }
\vbox{
\psfig{figure=f11b.eps,height=6cm,angle=270,width=8cm}
\psfig{figure=f11c.eps,height=6cm,angle=270,width=8cm}
\psfig{figure=f11d.eps,height=6cm,angle=270,width=8cm}}
\vbox{
\psfig{figure=f11e.eps,height=6cm,angle=270,width=8cm}
\psfig{figure=f11f.eps,height=6cm,angle=270,width=8cm}
\psfig{figure=f11g.eps,height=6cm,angle=270,width=8cm} }}

\caption{Comparison between the derived spectral parameters and
the sources' properties (see \S\ref{discussion} for details). To infer
the 1--10\,keV luminosity we assumed a distance of 3, 3, 5, 3.3, 3, 7,
10, 4, 5, and 10\,kpc, for the sources ordered as the labels reported
in the top panel (from left to right and top to bottom). Errors in the
luminosities are assumed to be 30\% of the reported values (which is
of the order of the flux errors), although the real error (including
that on the distance) is actually much larger.}}
\label{correlations}
\end{figure*}


\label{lastpage}


\begin{thebibliography}{}


\bibitem {}
   Anders, E. \& Grevesse, N., 1989, Geochimica \& Cosmochimica Acta 53, 197
\bibitem {}
Balucinska-Church, M. \& McCammon, D. 1992, ApJ, 400, 699
\bibitem {}
   Beloborodov, A. M. \&  Thompson, C.  2007, ApJ, 657, 967
\bibitem {}
Burgay, M., Rea, N.,  Israel, G. L., Possenti, A.,  Burderi, L., di Salvo, T., D'Amico, N.,  Stella, L. 2006, MNRAS, 372, 410
\bibitem {}
Burgay, M., Rea, N., Israel, G. L., Possenti, A. 2007, ATel, \# 903
\bibitem {}
Campana, S.,  Rea, N., Israel, G. L., Turolla, R., Zane, S., 2007, A\&A, 463, 1047
\bibitem {}
Chatterjee, P., Hernquist, L., \& Narayan, R., 2000, ApJ, 534, 373
\bibitem {}
        Camilo, F., Ransom, S. M., Halpern, J. P., Reynolds, J., Helfand, D. J.; Zimmerman, N., Sarkissian, J. 2006, Nature, 442, 892
\bibitem {}
Camilo, F., Ransom, S. M., Halpern, J. P., Reynolds, J. 2007a, ApJ, 666, L93
\bibitem {}
Camilo, F. \& Reynolds, J., 2007b, ATel \# 1056
\bibitem {}
Dall'Osso, S., Israel, G.L., Stella, L., Possenti, A., \& Perozzi, E. 2003, ApJ, 499, 485
\bibitem {}
Dib, R., Kaspi, V. M., Gavriil, F.  2007a, ApJ in press, arXiv0706.4156
\bibitem {}
Dib, R., Kaspi, V. M., Gavriil, F. P., Woods, P. M. 2007b, ATel \# 104
\bibitem {}
     Duncan, R.C., \& Thompson, C. 1992, ApJ, 392, L9
\bibitem {}
       Durant, M., \& van Kerkwijk, M. H., 2006, ApJ, 650, 1082
\bibitem {}
       Fernandez, R., \& Thompson, C., 2007, ApJ, 660, 615
\bibitem {}
         Gavriil, F.P. \& Kaspi, V.M. 2002, ApJ, 567, 1067
\bibitem {}
         Gavriil, F.P., Kaspi V.M., \& Woods, P.M. 2002, Nature, 419, 142
\bibitem {}
         Gavriil, F.P. \& Kaspi, V.M. 2004, ApJ, 609, L67
\bibitem {}
Gavriil, Fotis P., Kaspi, V. M., \& Woods, P. M. 2006, ApJ, 641, 418
\bibitem {}
Gelfand, J. D. \& Gaensler, B. M., 2007, ApJ, 667, 1111
\bibitem {}
Goldreich, P. \& Julian W.H. 1969, ApJ, 157, 869
\bibitem {}
Gonzalez, M. E., Dib, R., Kaspi, V. M., Woods, P. M., Tam, C. R., Gavriil, F. P 2007, ApJ submitted (arXiv0708.2756)
\bibitem {}
G\"otz, D., Mereghetti, S.,  Tiengo, A.,  Esposito, P. 2006, A\&A, 449, L31
\bibitem {}
G\"otz, D.,  et al. 2007, A\&A, 475, 317
\bibitem {}
Gotthelf, E. V.; Halpern, J. P.; Buxton, M.; Bailyn, C, 2004, ApJ, 605, 368
\bibitem{}
G\"uver, T., \"Ozel, F., G\"o\"g\"us, E., Kouveliotou, C., 2007a, ApJ, 667, L73
\bibitem{}
G\"uver, T., \"Ozel, F., \& G\"o\"g\"us, 2007b, ApJ in press (arXiv0705.3982)
\bibitem {}
Haberl, F., Freyberg, M. J., Briel, U. G., Dennerl, K., Zavlin, V. E. 2004, SPIE, 5165, 104
\bibitem {}
         Haberl, F., Zavlin, V.E.,  Trumper, J. \& Burwitz, V., 2004, A\&A 419, 1077
\bibitem {}
Haberl, F. 2007, Ap\&SS, 308, 73
\bibitem {}
Halpern, J. P., Gotthelf, E. V., Reynolds, J., Ransom, S. M., Camilo, F., 2007, ApJ in press (arXiv0711.3780)
\bibitem {}
Ho, W.C.G. \&  Lai, D. 2003, MNRAS, 338, 233
\bibitem {}
   Hulleman, F., van Kerkwijk, M.H. \& Kulkarni, S.R. 2000, Nature, 408, 689
\bibitem {}
Hurley, K., et al. 2005, Nature, 434, 1098
\bibitem {}
Ibrahim, A. I., et al. 2004, ApJ, 609, L21
  \bibitem {}
         Israel, G. L., G\"otz, D., Zane, S., Dall'Osso, S., Rea, N., Stella, L. 2007a, A\&A, 476, L9I
   \bibitem {}
         Israel, G. L., Campana, S., Dall'Osso, S., Muno, M. P., Cummings, J., Perna, R., Stella, L. 2007b, ApJ, 664, 448
\bibitem[2001]{j:01}
        Jansen, F., et al. 2001, A\&A, 365, L1
\bibitem {}
Kaspi, V.M., Gavriil, F.P., Woods, P.M., Jensen, J.B., Roberts, M.S.E., \& Chakrabarty, D., 2003, ApJ, 588, L93
\bibitem {}
        Kuiper,L., Hermsen, W.,  \& M\'endez, M. 2004, ApJ, 613, 1173
\bibitem {}
        Kuiper, L., Hermsen, W.,  den Hartog, P. R., Collmar, W., 2006, ApJ, 645, 556
\bibitem {}
Lamb, D. Q., Wang, J. C. L. \& Wasserman, I. 1990, ApJ, 363, 670
\bibitem {}
        Lyutikov, M., \& Gavriil F.P., 2006, MNRAS, 368, 690
\bibitem[\protect\citeauthoryear{Lebrun et al.}{2003}]{isgri}
 Lebrun, F., Leray, J.P., Lavocat, P., et al. 2003, A\&A, 411, L141
\bibitem {}
Lodders, K. 2003, ApJ, 591, 1220
\bibitem {}
McLaughlin, M. A., et al. 2006, Nature, 439, 817
\bibitem {}
Marsden, D. \&  White, N. E. 2001, ApJ, 551, L155
\bibitem {}
Mereghetti, S., Tiengo, A., Stella, L., Israel, G.L., Rea, N., Zane, S., \& Oosterbroek, T., 2004, ApJ, 608, 427
\bibitem {}
Mereghetti, S., et al. 2005, ApJ, 628, 938
\bibitem {}
Mereghetti, S., G\"otz, D., Mirabel, I. F., Hurley, K. 2005, A\&A 433, L9
\bibitem[\protect\citeauthoryear{Molkov et al.}{2005}]{molkov}
\bibitem {}
Mereghetti, S., 2008, A\&AR submitted (arXiv:0804.0250)
\bibitem {}
Molkov, S.,  Hurley, K., Sunyaev, R., et al., 2005, A\&A, 433, L13
\bibitem {}
Muno, M. P., Gaensler, B. M., Clark, J. S., de Grijs, R., Pooley, D., Stevens, I. R., Portegies Zwart, S. F 2007, MNRAS, 378, L44

\bibitem {}
Nagel, W. 1981, ApJ, 251, 288
\bibitem {}
Nobili, L., Turolla, R. \& Zampieri, L. 1993, ApJ, 404, 686
\bibitem {}
Nobili, L., Turolla, R. \& Zane, S., 2008, MNRAS in press, arXiv:0802.2647
\bibitem {}
Palmer, D. M.,  et al. 2005, Nature, 434, 1107
\bibitem {}
        Perna, R., Heyl, J., Hernquist, L., 2000, ApJ, 541, 344
\bibitem[2005]{r:03}
        Rea, N.,  et al. 2005, MNRAS, 361, 710
\bibitem {}
      Rea, N., Turolla, R., Zane, S., Tramacere, A., Israel, G.L., Stella, L., Campana, R., 2007a, ApJ, 661, L65
\bibitem {}
      Rea, N., Zane, S., Lyutikov, M. \& Turolla, R., 2007b, Ap\&SS, 308, 61
\bibitem {}
   Rea, N.,  et al.  2007c,  MNRAS, 381, 293
\bibitem[2001]{st:01}
        Str\"uder, L. et al. 2001, A\&A, 365, L18
\bibitem {}
Tam, C. R., Gavriil, F.,  Dib, R., Kaspi, V. M.; Woods, P. M., Bassa, C. 2007, ApJ in press (arXiv0707.2093)
\bibitem[\protect\citeauthoryear{Thompson \& Belobodorov}{2005}]{tb05}
Thompson, C., \& Beloborodov, A.~M., ApJ  634, 565 (2005)
\bibitem {}
        Thompson, C., \& Duncan, R.C., 1993, ApJ, 408, 194
\bibitem {}
        Thompson, C., \& Duncan, R.C. 1995, MNRAS, 275, 255
\bibitem {}
        Thompson, C., \& Duncan, R.C. 1996, ApJ, 473, 322
\bibitem {}
 Thompson, C., Lyutikov, M. \& Kulkarni, S.R., 2002, ApJ, 574, 332
\bibitem[2001]{tu:01}
        Turner, M. J. L. et al. 2001, A\&A, 365, L27
\bibitem {}
Turolla, R., Zane, S. \& Drake, J.J. 2004, ApJ, 603, 265
\bibitem[\protect\citeauthoryear{Ubertini et al.}{2003}]{ibis}
Ubertini, P.,  et al. 2003, A\&A 411, L131
\bibitem {}
    van Adelsberg, M. \& Lai, D. 2006, MNRAS, 373, 1495
\bibitem {}
        van Paradijs, J., Taam, R.E., \& van den Heuvel, E.P.J., 1995, A\&A, 299, L41
\bibitem {}
       Wasserman, I. \& Salpenter, E. 1980, ApJ, 498, 373
\bibitem {}
       Woods, P.M., et al. 2004, ApJ, 605, 378
\bibitem {}
       Woods, P.M., et al. 2005, ApJ, 629, 985
\bibitem {}
       Woods, P. \& Thompson, C., 2006, "Compact Stellar X-ray Sources", eds. W. H. G. Lewin \& M. van der Klis, 547
\end{thebibliography}
\end{document}